\documentclass[useAMS,usenatbib,iop,numberedappendix]{mn2e} 

\usepackage{mathptmx}
\usepackage{amsmath,amsfonts,amssymb}
\usepackage{makeidx}
\usepackage{graphicx}
\usepackage{epstopdf}
\usepackage{multirow}
\usepackage{rotating}
\usepackage{color}
\usepackage{tablefootnote}
\usepackage{lscape}
\usepackage{subfigure}
\usepackage{tablefootnote}
\usepackage[T1]{fontenc}
\usepackage{aecompl}

\definecolor{ored}{rgb}{1.00,0.27,0.00}
\definecolor{joerg}{rgb}{0.7, 0.4, 0.0}

\newcommand{\Munich}{$^{1}$}
\newcommand{\ExcellenceCluster}{$^{2}$}
\newcommand{\MPE}{$^{3}$}
\newcommand{\AIfA}{$^{4}$}
\newcommand{\FNAL}{$^{5}$}
\newcommand{\KICPChicago}{$^{6}$}
\newcommand{\AAUChicago}{$^{7}$}
\newcommand{\PhysicsUChicago}{$^{8}$}
\newcommand{\ANL}{$^{9}$}
\newcommand{\HarvardSmithsonian}{$^{10}$}
\newcommand{\UHarvard}{$^{11}$}
\newcommand{\MIT}{$^{12}$}
\newcommand{\UFlorida}{$^{13}$}
\newcommand{\Leiden}{$^{14}$}
\newcommand{\KIPACstandford}{$^{15}$}
\newcommand{\Ustanford}{$^{16}$}
\newcommand{\DarkCosmologyCenter}{$^{17}$}
\newcommand{\Melbourne}{$^{18}$}
\newcommand{\CTIO}{$^{19}$}

\author[Chiu et al.]{
I.~Chiu\Munich$^,$\ExcellenceCluster,
J.~P.~Dietrich\Munich$^,$\ExcellenceCluster,
J.~Mohr\Munich$^,$\ExcellenceCluster$^,$\MPE,
D.~E.~Applegate\AIfA,
B.~A.~Benson\FNAL$^,$\KICPChicago$^,$\AAUChicago,
\newauthor
L.~E.~Bleem\KICPChicago$^,$\PhysicsUChicago$^,$\ANL,
M.~B.~Bayliss\HarvardSmithsonian$^,$\UHarvard,
S.~Bocquet\Munich$^,$\ExcellenceCluster,
J.~E.~Carlstrom\KICPChicago$^,$\AAUChicago,
R.~Capasso\Munich$^,$\ExcellenceCluster,
\newauthor
S.~Desai\Munich$^,$\ExcellenceCluster,
C.~Gangkofner\Munich$^,$\ExcellenceCluster,
A.~H.~Gonzalez\UFlorida,
N.~Gupta\Munich$^,$\ExcellenceCluster,
C.~Hennig\Munich$^,$\ExcellenceCluster,
\newauthor
H.~Hoekstra\Leiden,
A.~von~der~Linden\KIPACstandford$^,$\Ustanford$^,$\DarkCosmologyCenter,
J.~Liu\Munich$^,$\ExcellenceCluster,
M.~McDonald\MIT,
C.~L.~Reichardt\Melbourne, 
\newauthor
A.~Saro\Munich$^,$\ExcellenceCluster,
T.~Schrabback\AIfA,
V.~Strazzullo\Munich,
C.~W.~Stubbs\HarvardSmithsonian$^,$\UHarvard,
A.~Zenteno\CTIO
\\
\Munich Faculty of Physics, Ludwig-Maximilians University, Scheinerstr.\ 1, 81679 Munich, Germany \\
\ExcellenceCluster Excellence Cluster Universe, Boltzmannstr.\ 2, 85748 Garching, Germany \\
\MPE Max Planck Institute for Extraterrestrial Physics, Giessenbachstr.\ 85748 Garching, Germany \\
\AIfA Argelander-Institut f\"ur Astronomie, Auf dem H\"ugel 71, D-53121 Bonn, Germany \\
\FNAL Fermi National Accelerator Laboratory, Batavia, IL 60510-0500 \\
\KICPChicago Kavli Institute for Cosmological Physics, University of Chicago, 5640 South Ellis Avenue, Chicago, IL 60637 \\
\AAUChicago Department of Astronomy and Astrophysics, University of Chicago, 5640 South Ellis Avenue, Chicago, IL 60637 \\
\PhysicsUChicago Department of Physics, University of Chicago, 5640 South Ellis Avenue, Chicago, IL 60637 \\
\ANL Argonne National Laboratory, 9700 S. Cass Avenue, Argonne, IL, USA 60439 \\
\HarvardSmithsonian Harvard-Smithsonian Center for Astrophysics, 60 Garden
Street, Cambridge, MA 02138 \\
\UHarvard Department of Physics, Harvard University, 17 Oxford Street,
Cambridge, MA 02138 \\
\UFlorida Department of Astronomy, University of Florida, Gainesville, FL 32611 \\
\Leiden Leiden Observatory, Leiden University, PO Box 9513, 2300 RA, Leiden, the Netherlands \\
\KIPACstandford Kavli Institute for Particle Astrophysics and Cosmology (KIPAC), Stanford University, 452 Lomita Mall, Stanford, CA 94305-4085, USA \\
\Ustanford Department of Physics, Stanford University, 452 Lomita Mall, Stanford, CA 94305-4085, USA \\
\DarkCosmologyCenter Dark Cosmology Centre, Niels Bohr Institute, University of Copenhagen Juliane Maries Vej 30, 2100 Copenhagen, Denmark \\
\MIT Kavli Institute for Astrophysics and Space Research, Massachusetts Institute of Technology, 77 Massachusetts Avenue, Cambridge, MA 02139 \\
\Melbourne School of Physics, University of Melbourne, Parkville, VIC 3010, Australia \\
\CTIO Cerro Tololo Inter-American Observatory, Casilla 603, La Serena, Chile
}

\newcommand{\LCDM}{\ensuremath{\Lambda\textrm{CDM}}}
\newcommand{\OmegaM}{\ensuremath{\Omega_{\mathrm{M}}}}
\newcommand{\OmegaL}{\ensuremath{\Omega_{\Lambda} } }

\newcommand{\Hnow}{\ensuremath{H_{0}}}

\newcommand{\Msun}{\ensuremath{\mathrm{M}_{\odot}}}
\newcommand{\Rfiveoo}{\ensuremath{R_{500}}}
\newcommand{\Rfiveoosz}{\ensuremath{R_{500-\mathrm{SZE}}}}
\newcommand{\Mfiveoo}{\ensuremath{M_{500}}}

\newcommand{\Mfiveoosz}{\ensuremath{M_{500-\mathrm{SZE}}}}

\newcommand{\massfactor}{\ensuremath{\eta}}
\newcommand{\redshift}{\ensuremath{z}}

\newcommand{\zd}{\ensuremath{z_\mathrm{l}}}
\newcommand{\Dd}{\ensuremath{D_\mathrm{l}}}
\newcommand{\Ds}{\ensuremath{D_\mathrm{s}}}
\newcommand{\Dds}{\ensuremath{D_\mathrm{ls}}}
\newcommand{\dif}{\ensuremath{\mathrm{d}}}

\newcommand{\bvec}{\ensuremath{\boldsymbol}}
\newcommand{\grad}{\ensuremath{\nabla}}
\newcommand{\convg}{\ensuremath{\kappa}}
\newcommand{\shear}{\ensuremath{\gamma}}
\newcommand{\sheartan}{\ensuremath{g_{+}}}
\newcommand{\shearcro}{\ensuremath{g_{\times}}}
\newcommand{\magni}{\ensuremath{\mu}}
\newcommand{\mcut}{\ensuremath{m_{\mathrm{cut}}}}

\newcommand{\gr}{\ensuremath{g-r}}
\newcommand{\ri}{\ensuremath{r-i}}
\newcommand{\uband}{\ensuremath{u}}
\newcommand{\gband}{\ensuremath{g}}
\newcommand{\rband}{\ensuremath{r}}
\newcommand{\iband}{\ensuremath{i}}
\newcommand{\zband}{\ensuremath{z}}

\newcommand{\lensingeff}{\ensuremath{\left<\beta\right>}}
\newcommand{\fends}{\ensuremath{s}}
\newcommand{\Pz}{\ensuremath{P(z)}}
\newcommand{\Ptz}{\ensuremath{P_{\mathrm{t}}(z)}}
\newcommand{\Pb}{\ensuremath{P(\beta)}}

\newcommand{\nd}{\ensuremath{n^\mathrm{}}}
\newcommand{\Nd}{\ensuremath{N^\mathrm{}}}
\newcommand{\Area}{\ensuremath{A_{\mathrm{ann}}}}
\newcommand{\nzero}{\ensuremath{n_{0}}}
\newcommand{\Ntotmodel}{\ensuremath{N_{\mathrm{mod}}}}
\newcommand{\nmodel}{\ensuremath{n_{\mathrm{mod}}}}
\newcommand{\Ntot}{\ensuremath{N_{\mathrm{tot}}}}
\newcommand{\Ncl}{\ensuremath{N_{\mathrm{cl}}}}
\newcommand{\Cstat}{\ensuremath{C_{\mathrm{stat}}}}
\newcommand{\Nradbin}{\ensuremath{N_{\mathrm{bins}}}}

\newcommand{\magninety}{\ensuremath{m_{90}}}
\newcommand{\magfifty}{\ensuremath{m_{50}}}
\newcommand{\magsig}{\ensuremath{\sigma_{\mathrm{m}}}}
\newcommand{\fmask}{\ensuremath{f_\mathrm{umsk}}}
\newcommand{\fcom}{\ensuremath{f_\mathrm{com}}}
\newcommand{\percent}{\ensuremath{\%}}

\DeclareMathOperator\erf{erf}

\newcommand{\mfitlow}{\ensuremath{1.30}}
\newcommand{\mfitlowonesigma}{\ensuremath{^{+0.41}_{-0.39}}}
\newcommand{\mfitlowsys}{\ensuremath{\pm0.13}}

\newcommand{\mfithi}{\ensuremath{0.46}}
\newcommand{\mfithionesigma}{\ensuremath{^{+0.33}_{-0.29}}}
\newcommand{\mfithisys}{\ensuremath{\pm0.036}}

\newcommand{\mfitcomb}{\ensuremath{0.83}}
\newcommand{\mfitcombonesigma}{\ensuremath{\pm0.24}}
\newcommand{\mfitcombsys}{\ensuremath{\pm0.074}}

\newcommand{\syssig}{\ensuremath{\sigma_{\mathrm{sys}}^{\mathrm{tot}}}}

\title[Magnification bias of background galaxies]{Detection of Enhancement in Number Densities of Background Galaxies due to Magnification by Massive Galaxy Clusters}

\begin{document}
%set A4 size
\pdfpageheight 11.7in
\pdfpagewidth 8.3in
%\date{Accepted ???. Received ???; in original form ???}  

%%%%%%%%%%%%%%%%%%%%%%%%%%%%%%%%%%%%%%%%%%
%
% TITLE
%
%%%%%%%%%%%%%%%%%%%%%%%%%%%%%%%%%%%%%%%%%%

\maketitle 

%%%%%%%%%%%%%%%%%%%%%%%%%%%%%%%%%%%%%%%%%%
%
% ABSTRACT
%
%%%%%%%%%%%%%%%%%%%%%%%%%%%%%%%%%%%%%%%%%%

\begin{abstract}
We present a detection of the enhancement in the number densities of background galaxies induced from lensing magnification and use it to test the Sunyaev-Zel'dovich effect (SZE) inferred masses in a sample of 19 galaxy clusters with median redshift $\redshift\simeq0.42$ selected from the South Pole Telescope SPT-SZ survey.  
These clusters are observed by the Megacam on the Magellan Clay Telescope though $gri$ filters.
Two background galaxy populations are selected for this study through their photometric colours; they have median redshifts ${\redshift}_{\mathrm{median}}\simeq0.9$ (low-$z$ background) and ${\redshift}_{\mathrm{median}}\simeq1.8$ (high-$z$ background).  Stacking these populations, we detect the magnification bias effect at $3.3\sigma$ and $1.3\sigma$ for the low- and high-\redshift\ backgrounds, respectively.  We fit NFW models simultaneously to all observed magnification bias profiles to estimate the multiplicative factor \massfactor\ that describes the ratio of the weak lensing mass to the mass inferred from the SZE observable-mass relation. We further quantify systematic uncertainties in \massfactor\ resulting from the photometric noise and bias, the cluster galaxy contamination and the estimations of the background properties.  
The resulting \massfactor\ for the combined background populations with $1\sigma$ uncertainties is 
$\mfitcomb\mfitcombonesigma\mathrm{(stat)}\mfitcombsys\mathrm{(sys)}$,  indicating good consistency between the lensing and the SZE-inferred masses.  We use our best-fit \massfactor\ to predict the weak lensing shear profiles and compare these predictions with observations, showing agreement between the magnification and shear mass constraints.  This work demonstrates the promise of using the magnification as a complementary method to estimate cluster masses in large surveys.
\end{abstract}

%%%%%%%%%%%%%%%%%%%%%%%%%%%%%%%%%%%%%%%%%%
%
% KEYWORDS
%
%%%%%%%%%%%%%%%%%%%%%%%%%%%%%%%%%%%%%%%%%%


%%%%%%%%%%%%%%%%%%%%%%%%%%%%%%%%%%%%%%%%%%
%
% INTRODUCTION
%
%%%%%%%%%%%%%%%%%%%%%%%%%%%%%%%%%%%%%%%%%%

\section{Introduction}
\label{sec:introduction}

Gravitational lensing is one of the most direct methods for
measuring the masses of galaxy clusters, because it does not require
assumptions about the dynamical or hydrostatic state of the clusters 
and it probes the total underlying mass distribution. 
In practice, there are challenging observational
systematics that must be overcome \citep{erben01,leauthaud07,corless09,viola11,hoekstra13}, and over the past two decades
significant progress has been made by calibrating with simulations
\citep[e.g.,][]{heymans06,massey07,bridle09,kitching12,mandelbaum14b,hoekstra15}.
As a result, modelling the shear distortion of background galaxies that are lensed
has been developed into a reliable method to measure
cluster masses 
\citep{
gruen14,umetsu14,
applegate14,vonderlinden14a,
vonderlinden14b,hoekstra15}.
In comparison, there has until recently
been less observational progress using the complementary gravitational
lensing magnification effect
\citep{broadhurst95,dye2002,joachimi10,vanwaerbeke10,heavens10,
hildebrandt11,schmidt12,umetsu13,
coupon13,ford13,duncan14}.

The changes in the sizes of the background galaxy population due to
gravitational lensing magnification results in changes to the fluxes
because the surface brightness is conserved. This leads to increases
in the number density of flux-selected samples of background galaxies
in the neighborhood of mass concentrations. 
However, the magnification effect also distorts
the sky area, leading to a decrease in
the number density. Whether the combined effects lead to an overall
increase or decrease of the number density depends on the slope of the
source count-magnitude relation at the flux limit. An advantage to
measuring the magnification is that it only requires accurate
photometry and therefore does not require unbiased estimates of
galaxy ellipticity, which are needed for shear studies. 
Thus, even unresolved galaxy populations can be used in a lensing
magnification study. However, the signal-to-noise ratio (SNR) for mass
measurements obtained using magnification effects tends to be lower by
a factor of 3--5 as compared to those that one obtains using the shear
signature imprinted on the same galaxies \citep{schneider00}. Due to
the lower SNR, a significant detection of the magnification effect is
more realistically expected around massive collapsed structures such
as galaxy clusters.

There are several ways to detect the magnification around galaxy
clusters. The magnification information can be extracted from the angular cross-correlation
of high redshift sources, e.g., Lyman
break galaxies \citep{hildebrandt09,
  vanwaerbeke10,hildebrandt11,ford12, ford13}, measuring the change in
the background galaxy sizes or fluxes \citep{schmidt12}, 
simultaneously estimating the increase in the observed number counts and fluxes of the background luminous red galaxies \citep{bauer14},
or observing
the skewness in the redshift distribution of the background galaxies
\citep{coupon13,jimeno15}. Another approach, called the \emph{magnification
  bias}, is to measure the change or bias in the number density of a
flux-limited background galaxy sample towards the cluster centre
\citep{broadhurst95, taylor98}. First proposed by \cite{broadhurst95},
who measured the mass of an individual cluster with this technique,
the magnification bias method has now been applied to a dozen galaxy
clusters \citep{umetsu13}. In that analysis, the magnification bias
signature is combined not only with shear but also with strong lensing
constraints.

The conventional analysis of magnification bias is based on a
flux-limited background galaxy population with a nearly flat slope of
the source count-magnitude relation, which leads to a depletion of the
number density in the mass-concentrated region of clusters
\citep{umetsu13}. Detecting this magnification bias requires
ultra-deep and uniform observations to achieve adequate statistics in
the galaxy counts to suppress the Poisson noise. Therefore, this
approach for measuring the cluster masses can be very costly in terms
of observing time. On the other hand, the lensing magnification also
acts on brighter galaxies where the intrinsic slope is steep. In this
case, the increase of the number of galaxies magnified to be above the
flux limit overcomes the dilution of the geometric expansion and,
therefore, results in an enhancement of number density. However, this
density enhancement of the magnification bias has a lower SNR on a per
cluster basis due to the lower number density of bright background
galaxies. Consequently, one needs to combine the signal from a large
sample of massive clusters.

In this work, we aim to detect the density enhancement from the
magnification bias effect by combining information from 19 massive
clusters. Our study leverages background populations of normal
galaxies selected in colour-colour space. The clusters were selected
through their Sunyaev-Zel'dovich effect
\citep[SZE;][]{sunyaev70,sunyaev72} in the 2500~deg$^2$ SPT-SZ survey
carried out using the South Pole Telescope
\citep[SPT,][]{carlstrom11}. These clusters have been subsequently imaged
with the Magellan telescope for the purpose of weak lensing studies.
It is worth mentioning that our approach is similar to the number count method conducted in \cite{bauer14} with the difference that they only used the background populations of the luminous red galaxies with $i$-band magnitude brighter than $\approx20$~mag, while in this work we extend the background samples to the normal galaxies at much fainter limiting magnitudes.

This paper is organized as follows: A brief review of the relevant
lensing theory is given in Section~\ref{sec:theory}. In
Section~\ref{sec:sample_and_data} we introduce the data used for this
analysis. The analysis method is described in detail in
Section~\ref{sec:analysis}. We present and discuss our results in
Section~\ref{sec:results_and_discussion} and provide our conclusions
in Section~\ref{sec:conclusion}. Throughout this paper, we assume the
concordance \LCDM\ cosmological model with the cosmological parameter
values recently determined by \cite{bocquet14}: \OmegaM = 0.292,
\OmegaL = 0.708 and \Hnow\ = 68.2~km~s$^{-1}$ Mpc$^{-1}$. Unless
otherwise stated, all uncertainties are 68\percent\ ($1\sigma$) confidence intervals and
cluster masses and radii are estimated within a region that has an
overdensity of 500 with respect to the critical density of the
Universe at the cluster redshift.
The magnitudes in this work are all in the AB magnitude system.
The distances quoted in this work are all in physical units.

%%%%%%%%%%%%%%%%%%%%%%%%%%%%%%%%%%%%%%%%%%%%%%
%
% Theory
%
%%%%%%%%%%%%%%%%%%%%%%%%%%%%%%%%%%%%%%%%%%%%%%

\section{Theory}
\label{sec:theory}

In this section we provide a summary of gravitational lensing induced
by galaxy clusters. We refer the reader to \cite{umetsu10} and
\cite{hoekstra13} for more complete discussions.

Light traveling from a distant source to the observer is deflected
in the presence of a gravitational potential, resulting in the
distortion of the observed image. This gravitational lensing effect
depends only on the underlying mass distribution along the line of
sight and can be formulated with the following lens equation:
\begin{equation}
\label{eq:lensing_equation}
\bvec{\alpha} = \bvec{\theta} - \bvec{\grad_{\theta}}{\psi}  \ ,
\end{equation}
where $\psi$ is the effective deflection potential, 
$\bvec{\alpha}$ and $\bvec{\theta}$ are the angular positions on the
sky of the source (before lensing) and the observed image (after
lensing), respectively. The Jacobian of
equation~(\ref{eq:lensing_equation}) therefore reflects how the
observed background image is distorted, linking the positions of the
source and the gravitational potential of the lens. i.e.,
\begin{eqnarray}
\label{eq:jacobian}
\mathbfss{J}(\bvec{\theta}) &= &\grad_{\bvec{\theta}}\bvec{\alpha} \nonumber \\
 &= &\left( 
 \begin{array}{cc}
 1 - \convg -\shear_1    &-\shear_2 \\
 -\shear_2  &1 - \convg +\shear_1 
 \end{array}
\right)   
\end{eqnarray}   
and
\begin{equation}
\label{eq:area_distortion_def}
\dif\Omega_{\bvec{\theta}} = \mathbfss{J}^{-1}\dif\Omega_{\bvec{\alpha}} \ ,
\end{equation}
where $\convg$ and $\bvec{\shear} = \shear_1+ i\shear_2$ are,
respectively, the convergence and the shear at the sky position of the
image; $\dif \Omega_{\bvec{\alpha}}$ and $\dif \Omega_{\bvec{\theta}}$
denote the solid angle on the sky before and after lensing,
respectively. The convergence \convg\ is the integrated density
contrast against the background along the line of sight. 
For the case of cluster lensing, \convg\ can
be written as
\begin{eqnarray}
\label{eq:kappa_def}
\convg(\bvec{\theta}, \psi)	 & =  &\frac{\Sigma_{\mathrm{lens}}(\bvec{\theta}, \psi)}{\Sigma_{\mathrm{crit}} } \, , \\
\label{eq:sigmam_def}
\Sigma_{ \mathrm{crit} }	 &=  &
\frac{c^2}{4 \upi G} \frac{1}{ \beta \Dd} \, , \; \text{and} \\
\label{eq:lensing_eff_def}
\beta &= &\left\lbrace
\begin{array}{cr}
0     &\mathrm{for} \; \Ds \le \Dd \\
\frac{\Dds}{ \Ds } &\mathrm{for} \; \Ds > \Dd 
\end{array}
\right.
\end{eqnarray}
assuming that the cluster acts as a single thin lens 
ignoring the uncorrelated large-scale structure, 
i.e., an
instantaneous deflection of the light ray. Here
${\Sigma}_{\mathrm{lens}}$ is the projected mass density of the
cluster, ${\Sigma}_{\mathrm{crit}}$ is the critical surface mass
density, $\beta$ is the lensing efficiency that depends on the ratio
of the lens-source distance to the source distance averaged over the
population of background galaxies, $c$ is the speed of light, and \Dd,
\Ds\ and \Dds\ denote the angular diameter distances of the cluster,
the source, and between the cluster and the source,
respectively. These distances depend on the observed redshifts and the
adopted cosmological parameters.  
In practice, the lensing efficiency averaged over a population \lensingeff\ is used for estimating cluster masses.

As seen from equation~(\ref{eq:jacobian}), gravitational lensing
induces two kinds of changes to the observed image. The first one,
characterized by $\bvec{\shear}$, distorts the observed image
anisotropically, while the other described by the convergence
\convg\ results in an isotropic magnification. Analyzing the
information from shear alone can only recover the gradient of the
cluster potential, and therefore the inferred mass is subject to an
arbitrary mass constant. This so-called mass-sheet degeneracy
can be broken by combining shear and magnification
\citep[e.g.][]{seitz97}.

As seen in equation~(\ref{eq:area_distortion_def}), gravitational
lensing changes the projected area of the observed image, and because
the surface brightness is conserved this results in a magnification
\magni\ of the source, which is given by
\begin{eqnarray}
\label{eq:magni_def}
\magni & = &\mathrm{det}(\mathbfss{J})^{-1} \nonumber \\
	&= &\frac{1}{ (1-\convg)^2 - \|\shear\|^2} \, .
\end{eqnarray}
In the weak lensing limit ($\|\bvec{\shear}\| \ll 1$ and $\convg \ll 1$),
the magnification can be approximated as $\magni \simeq 1+2\convg$,
i.e. it is linearly related to the dimensionless surface mass density
\convg.
 
For $\magni>1$ the flux of each source is increased, leading to an
increase in the observed number density of a flux-limited population
of background sources. On the other hand, the lensing magnification
introduces an angular expansion on the plane of the sky, which
decreases the observed number of background sources per unit area. As
a result, the observed number density of a flux-limited background
population changes (is either depleted or enhanced) towards the centre
of the cluster depending on the two competing effects. The mass of a
cluster can hence be estimated by measuring this change given
knowledge of the properties of the observed background population
prior to lensing.

One important property of the background population is its number
count-magnitude relation $n(<m)$, which is the cumulative number of
galaxies per unit sky area brighter than a particular magnitude
$m$. This number count-flux relation is typically characterized as a
power law $n(<f)=f_0\times f^{-2.5s}$ where $f$ is flux, $f_0$ is a
normalization and $s$ is the power law index. This can be written in
terms of magnitude $m$ as
\begin{equation}
\label{eq:counts}	
\log n(<m)=\log f_0 + s\times (m - \mathrm{ZP}) \, ,
\end{equation} 
where $\mathrm{ZP}$ is the zeropoint used to convert the flux to magnitude.
In the presence of lensing the observed cumulative number density
\nd$(<\mcut)$ of a given background population can be shown to be
\citep{broadhurst95,umetsu11}
\begin{eqnarray}
\label{eq:observed_n}
\nd (< \mcut) &= &\nzero(<\mcut) \, \magni^{2.5\fends-1}  \\
& & \nonumber \\
\label{eq:fends_def}
\fends(\mcut) &= &\frac{\dif  \log n(<m)}{\dif m} 
\bigg|_{\mcut}  \ ,
\end{eqnarray}
where $\nzero(<\mcut)$ is the projected number density of galaxies at
the threshold magnitude \mcut\ in the absence of lensing and
$\fends(\mcut)$ is the power law index of the galaxy count-magnitude
distribution before lensing (equation~(\ref{eq:counts})) evaluated at
the limiting magnitude $\mcut$. Equation~(\ref{eq:observed_n}) can be
further reduced to
\begin{equation}
\nd (<m_\mathrm{\mbox{\scriptsize cut}}) \simeq \nzero(<\mcut) (1+(5\fends-2)\convg)\label{eq:nd2nzero}
\end{equation}
in the weak lensing regime.

In the case of $\fends\ = 0.4$, one expects no magnification signal while
a background population with \fends\ greater (less) than 0.4 results
in enhancement (depletion) of background objects. To sum up, the
cluster mass can be determined by using the magnification bias
information alone if the power law slope \fends, the average lensing
efficiency \lensingeff\ of the background population, and the local
background number counts \emph{before} lensing $\nzero(<\mcut)$ are
known.

%%%%%%%%%%%%%%%%%%%%%%%%%%%%%%%%%%%%%%%%%%
%
% SAMPLE AND DATA
%
%%%%%%%%%%%%%%%%%%%%%%%%%%%%%%%%%%%%%%%%%%

\section{Sample and Data}
\label{sec:sample_and_data}

\subsection{Sample}
\label{sec:sample}

We study the lensing magnification with 19 galaxy clusters selected by SPT through their SZE signatures. The first weak lensing shear based masses for five out of these 19 clusters have been presented in \cite{high12}, and the full sample is being examined in a subsequent weak lensing shear analysis (Dietrich et al, in preparation). These 19 clusters all have measured spectroscopic redshifts \citep{song12b,bleem15} and span the redshift range $0.28\le\redshift\le0.60$ with a median redshift of 0.42. The virial masses \Mfiveoo\ have been estimated using their SZE signature and the SZE mass-observable relation that has been calibrated using velocity
dispersions, X-ray mass proxies and through self-calibration in combination with external cosmological datasets that include Planck CMB anisotropy, WMAP CMB polarization anisotropy and SNe and BAO distances \citep{bocquet14}.

\begin{table*}
\centering
\caption{
Properties of the cluster sample.
Column~1: name. 
Column~2: spectroscopic redshift. 
Column~3--4: right ascension $\alpha_{2000}$ and declination $\delta_{2000}$ of the BCG. 
Column~5: the SZE-inferred \Mfiveoo\ (see Section~\ref{sec:sample}).
Column~6--7: \Rfiveoo\ corresponding to the SZE-inferred \Mfiveoo.
Column~8--10: 90\percent\ completeness limit (\magninety) for \gband, \rband\ and \iband\ filters, respectively.
}
\label{tab:sample_and_data_properties}
\begin{tabular}{cccccccccc}
\hline
Cluster   &Redshift    &$\alpha_{2000}$ & $\delta_{2000}$   &$\Mfiveoo$                &\multicolumn{2}{c}{\Rfiveoo}  &$\magninety^{\gband}$ &$\magninety^{\rband}$ &$\magninety^{\iband}$ \\
      &                 &[deg]  &[deg] &[$10^{14}$\Msun] &[Mpc] &[arcmin] &[mag] &[mag] &[mag] \\
\hline
    SPT-CL~J0234$-$5831    &    0.415    &     $ 38.676189 $     &     $ -58.523644 $     &     $ 9.03 \pm 1.76 $     &     $ 1.30 $     &     $ 3.82 $     &     $ 23.91 $     &     $ 24.54 $     &     $ 23.07 $         \\        
    SPT-CL~J0240$-$5946    &    0.400    &     $ 40.159710 $     &     $ -59.763600 $     &     $ 6.38 \pm 1.31 $     &     $ 1.16 $     &     $ 3.50 $     &     $ 24.05 $     &     $ 24.63 $     &     $ 23.21 $         \\        
    SPT-CL~J0254$-$5857    &    0.438    &     $ 43.564592 $     &     $ -58.952993 $     &     $ 8.77 \pm 1.70 $     &     $ 1.27 $     &     $ 3.63 $     &     $ 23.83 $     &     $ 24.21 $     &     $ 22.63 $         \\        
    SPT-CL~J0307$-$6225    &    0.579    &     $ 46.819712 $     &     $ -62.446544 $     &     $ 5.89 \pm 1.21 $     &     $ 1.05 $     &     $ 2.60 $     &     $ 24.24 $     &     $ 24.83 $     &     $ 23.58 $         \\        
    SPT-CL~J0317$-$5935    &    0.469    &     $ 49.315539 $     &     $ -59.591594 $     &     $ 4.71 \pm 1.11 $     &     $ 1.02 $     &     $ 2.81 $     &     $ 23.94 $     &     $ 24.54 $     &     $ 23.07 $         \\        
    SPT-CL~J0346$-$5439    &    0.530    &     $ 56.730934 $     &     $ -54.648699 $     &     $ 6.32 \pm 1.28 $     &     $ 1.10 $     &     $ 2.83 $     &     $ 24.26 $     &     $ 24.69 $     &     $ 23.47 $         \\        
    SPT-CL~J0348$-$4515    &    0.358    &     $ 57.071292 $     &     $ -45.250059 $     &     $ 7.04 \pm 1.41 $     &     $ 1.22 $     &     $ 3.94 $     &     $ 24.46 $     &     $ 25.13 $     &     $ 23.85 $         \\        
    SPT-CL~J0426$-$5455    &    0.630    &     $ 66.517205 $     &     $ -54.925319 $     &     $ 6.01 \pm 1.23 $     &     $ 1.04 $     &     $ 2.46 $     &     $ 24.13 $     &     $ 24.65 $     &     $ 23.21 $         \\        
    SPT-CL~J0509$-$5342    &    0.461    &     $ 77.339141 $     &     $ -53.703632 $     &     $ 5.87 \pm 1.21 $     &     $ 1.10 $     &     $ 3.06 $     &     $ 24.21 $     &     $ 24.59 $     &     $ 23.29 $         \\        
    SPT-CL~J0516$-$5430    &    0.295    &     $ 79.155613 $     &     $ -54.500493 $     &     $ 8.00 \pm 1.58 $     &     $ 1.30 $     &     $ 4.79 $     &     $ 23.41 $     &     $ 23.98 $     &     $ 22.64 $         \\        
    SPT-CL~J0551$-$5709    &    0.423    &     $ 87.898265 $     &     $ -57.141236 $     &     $ 5.77 \pm 1.20 $     &     $ 1.11 $     &     $ 3.24 $     &     $ 23.50 $     &     $ 24.06 $     &     $ 22.61 $         \\        
    SPT-CL~J2022$-$6323    &    0.383    &     $ 305.541020 $     &     $ -63.397044 $     &     $ 4.88 \pm 1.13 $     &     $ 1.07 $     &     $ 3.31 $     &     $ 23.68 $     &     $ 24.20 $     &     $ 22.56 $         \\        
    SPT-CL~J2030$-$5638    &    0.394    &     $ 307.688610 $     &     $ -56.632185 $     &     $ 4.12 \pm 1.10 $     &     $ 1.01 $     &     $ 3.06 $     &     $ 23.56 $     &     $ 24.09 $     &     $ 22.53 $         \\        
    SPT-CL~J2032$-$5627    &    0.284    &     $ 308.058670 $     &     $ -56.436827 $     &     $ 6.29 \pm 1.29 $     &     $ 1.21 $     &     $ 4.56 $     &     $ 23.26 $     &     $ 24.04 $     &     $ 22.22 $         \\        
    SPT-CL~J2135$-$5726    &    0.427    &     $ 323.914680 $     &     $ -57.437519 $     &     $ 7.02 \pm 1.39 $     &     $ 1.19 $     &     $ 3.44 $     &     $ 23.45 $     &     $ 23.96 $     &     $ 22.50 $         \\        
    SPT-CL~J2138$-$6008    &    0.319    &     $ 324.500020 $     &     $ -60.131848 $     &     $ 8.19 \pm 1.61 $     &     $ 1.30 $     &     $ 4.54 $     &     $ 22.92 $     &     $ 23.46 $     &     $ 21.71 $         \\        
    SPT-CL~J2145$-$5644    &    0.480    &     $ 326.466340 $     &     $ -56.748231 $     &     $ 7.85 \pm 1.53 $     &     $ 1.21 $     &     $ 3.27 $     &     $ 23.94 $     &     $ 24.37 $     &     $ 22.98 $         \\        
    SPT-CL~J2332$-$5358    &    0.402    &     $ 353.114480 $     &     $ -53.974436 $     &     $ 6.10 \pm 1.23 $     &     $ 1.14 $     &     $ 3.43 $     &     $ 24.26 $     &     $ 24.78 $     &     $ 23.66 $         \\        
    SPT-CL~J2355$-$5056    &    0.320    &     $ 358.947150 $     &     $ -50.927604 $     &     $ 4.80 \pm 1.10 $     &     $ 1.09 $     &     $ 3.79 $     &     $ 24.04 $     &     $ 24.78 $     &     $ 23.37 $         \\        
\hline
\end{tabular}
\end{table*}

Song et al. (2012) show that the Brightest Cluster Galaxy (BCG) position provides a good proxy for the cluster centre, which, for relaxed clusters, is statistically consistent with the centre inferred from the SZE map.
Moreover, the offset distribution between the BCG and SZE centres is consistent with the one between the BCG and X-ray centres that is seen in the local Universe \citep{lin04b}.  Therefore, the cluster centre is taken to be the
position of the BCG, which is visually identified on pseudo-colour images, in this work. \Rfiveoo\ is derived from the cluster SZE-inferred mass, its redshift and the critical density at that redshift, given the cosmological parameters. Properties of the 19 clusters are listed in Table~\ref{tab:sample_and_data_properties}.

\subsection{Data}
\label{sec:data}

The data acquisition, image reduction, source extraction, and the photometric calibration are described in \cite{high12}, to which we refer the reader for more details. In summary, the 19 galaxy clusters studied in this work were all observed using Megacam on the Magellan Clay 6.5-m telescope through $g^{\prime}$, $r^{\prime}$ and $i^{\prime}$ filters. The Megacam field of view is $25\arcmin\times25\arcmin$, which at the redshifts of our clusters covers a region around the cluster that extends to over $2.5\Rfiveoo$ and allows us to extract the background number density \nzero\ at large radii where the magnification effect is negligible. Except for
SPT-CL~J0516$-$5430, each cluster was observed through $g^{\prime}$ and $r^{\prime}$ filters in a three-point diagonal linear dither pattern with total exposure times of 1200~s and 1800~s, respectively, while a five-point diagonal linear dither pattern was used for $i^{\prime}$ band imaging with a total exposure time of 2400~s.
SPT-CL~J0516$-$5430 was observed with a $2\times2$ square dither mode and a total of eight pointings through the $g^{\prime}$, $r^{\prime}$ and $i^{\prime}$ filters with total exposure times of 1200~s, 1760~s,
and 3600~s, respectively.

Catalogs were created using \texttt{SExtractor}~\citep{bertin96} in dual image mode. Given that the $r^{\prime}$ images have the best seeing with the smallest variation, we use these as detection images. We adopt
\texttt{MAG\_AUTO} for photometry. The stellar locus together with 2MASS photometry is used both to determine zeropoint differences between bands \citep{high09} and the absolute zeropoint calibration \citep{song12b,desai12}. This results in the systematic uncertainties of colours $g^{\prime}-r^{\prime}$ and $r^{\prime}-i^{\prime}$ smaller than 0.03~mag. The absolute photometric calibration has uncertainties
of $\lesssim0.05$~mag.  Similarly to \cite{high12}, we convert our photometry from the SDSS system to the
Canada-France-Hawaii Telescope Legacy Survey (CFHTLS) system \citep{regnault09}\footnote{http://terapix.iap.fr/rubrique.php?id\_rubrique=241}.  
For convenience, we write $\gband$ instead of $\gband_{\mbox{\scriptsize CFHT}}$, and equivalently in other bands.

%%%%%%%%%%%%%%%%%%%%%%%%%%%%%%%%%%%%%%%%%%
%
% ANALYSIS
%
%%%%%%%%%%%%%%%%%%%%%%%%%%%%%%%%%%%%%%%%%%

\section{Analysis}
\label{sec:analysis}

We stack the galaxy count profiles of 19 clusters to enhance the SNR of the magnification bias and then fit a composite model that includes the individual cluster masking corrections, source count-magnitude distribution slope \fends\ and the lensing efficiency. This stacked analysis ends in a consistency test of the SZE inferred masses for the cluster ensemble. Details are provided in the subsections below.

%%%%%%%%%%%%%%%%
% BACKGROUND SELECTION
%%%%%%%%%%%%%%%%

\subsection{Source Catalog Completeness Limits}
\label{sec:completeness}

We estimate the completeness of the source catalog by comparing our number counts to that of a deep reference field where the source detection is complete in the magnitude range of interest in this work. 
In particular we extract the limiting magnitude where the completeness is 90\percent\ (\magninety) and 50\percent\ (\magfifty) for our source detection. Here we use the CFHTLS-DEEP survey \citep{ilbert06, coupon09}, in which the 80\percent\ completeness limits lie at magnitudes of $\uband=26.3$, $\gband=26.0$,
$\rband=25.6$, $\iband=25.4$ and $\zband=23.9$. Assuming that the complete source count-magnitude distribution can be described by a power law (i.e., $\log n(m)\propto a\times m + b$, where $a$ is the slope 
and $b$ is the normalization), we first derive its slope from the reference field using the magnitude range 20 to 24 in each band. Using this slope, we then fit the normalization of the source counts for galaxies brighter than 22~mag observed in the outskirts of our clusters ($r>2\Rfiveoo$). We use the ratio of the source counts in the cluster field to the derived best-fit power law to model the completeness function for each cluster as an error function. Specifically, the completeness function $F_{\mathrm{c}}$ is defined by
\begin{equation}
\label{eq:def_completeness_function}
F_{\mathrm{c}}(m) = \frac{1}{2} - \frac{1}{2}\erf\left(\frac{m -
    \magfifty}{\magsig}\right) \, , 
\end{equation}
where $\erf$ is the error function, \magfifty\ is the magnitude at which 50\percent\ completeness is reached, and \magsig\ is the characteristic width of the magnitude range over which the completeness decreases.

\begin{figure}
\centering
\resizebox{\hsize}{!}{\includegraphics[clip]{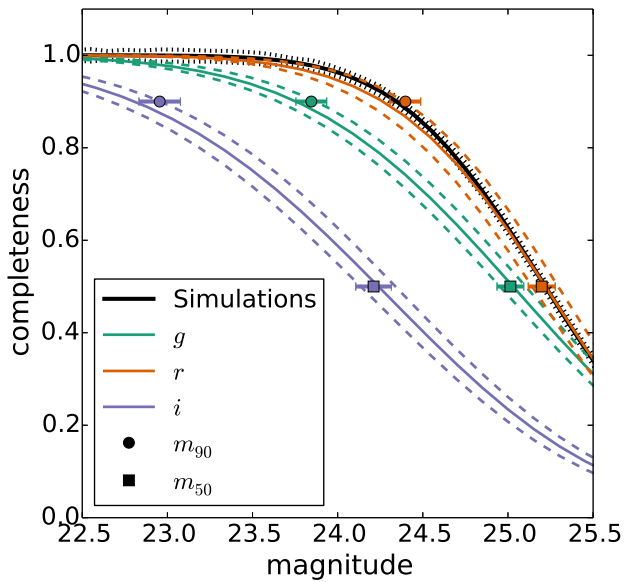}}
\vskip-0.2in
\caption{The completeness of the source detection as a function of  magnitude. The completeness derived from \gband, \rband\ and \iband\  source catalogs is plotted in the solid lines while the uncertainty of the mean is represented by the dashed lines. The solid circles and squares are the means of \magninety\ and \magfifty\ measured from the 19 cluster fields, respectively. Completeness functions for \gband, \rband\ and \iband\ are colour coded in green, orange and blue, respectively. The completeness function and its uncertainties measured on the mean of our image simulations are the black lines.
Note that the derived completeness is based on our catalogs obtained by running \texttt{SExtractor} in dual image mode with the \rband-band imaging as the detection band.
}
\label{fig:data_depth}
\end{figure}

We use the best-fit parameters of the completeness model for each cluster to derive the 90\percent\ completeness limit \magninety. We show the mean of the completeness functions as well as the measured \magninety\ and \magfifty\ of the 19 clusters for the three filters in Figure~\ref{fig:data_depth}. 

The mean \magninety\ of the 19 observed clusters is 23.84, 24.39 and 22.95 for the filters \gband, \rband\ and \iband, respectively. The \magninety's for the \gband, \rband\ and \iband\ passbands in each cluster are listed in Table~\ref{tab:sample_and_data_properties}.  Note that the depths in the \iband\  imaging limit our analysis at magnitudes fainter than 24~mag.  

After accounting for differences in primary mirror area, exposure time and quantum efficiency, we compare our completeness limits to those of SDSS Stripe 82 \citep{annis11}.   We estimate that in the background limited regime our Magellan imaging should be deeper by 1.1~mag, 1.2~mag and 1.3~mag in \gband\rband\iband, respectively, in comparison to SDSS Stripe 82.  Because the seeing is better in our Magellan imaging than in Stripe 82 we would expect these estimates to somewhat underestimate the true differences in the completeness limits.   A comparison of our 50\percent\ completeness limits \magfifty\ with theirs  (see Figure~7 in \cite{annis11}) indicates that our catalogs are deeper by $1.3\pm0.3$, $1.8\pm0.3$, $1.2\pm0.5$~mag, for \gband\rband\iband, respectively, indicating good consistency with expectation.  The comparison of \magninety\ in our two datasets leads to the same conclusion.

The source detection is also unavoidably affected by blending, especially in the crowded environment of clusters. We address how the blending affects the completeness of background galaxies with image simulations. With realistic image simulations we can quantify the incompleteness as a function of magnitude and distance from the
cluster centre and, therefore, apply a completeness correction to the analysis.

Specifically, we simulate images using \textsc{GalSim} \citep{rowe13} and derive the completeness of the sources detected by running \texttt{SExtractor} with the same configuration we use in the observed images.  We simulate 40 images with a set of galaxy populations and stars. Each image contains a spatially uniform distribution of background galaxies and foreground stars.

\begin{figure}
\centering
\resizebox{\hsize}{!}{\includegraphics[clip]{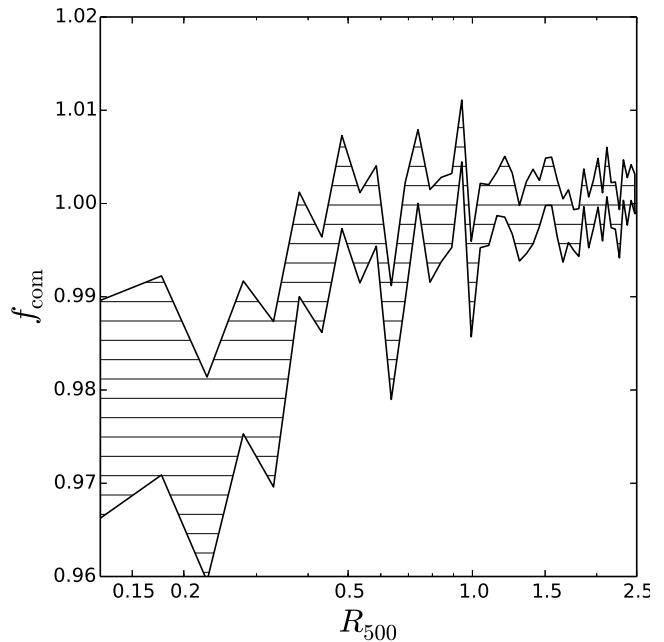}}
\caption{ The radial completeness \fcom(x)\ at $\mcut=23.5$ as a function of distance from the cluster centre derived from the simulations. The $1\sigma$ confidence region is filled with horizontal lines.}
\label{fig:fcom}
\end{figure}

We simulate background galaxies with a power law index $\fends=0.4$ of the source count-magnitude relation between the apparent magnitudes of 20 and 25.5 at $\redshift=0.9$, which is the median redshift of the low-\redshift\ background population studied here (see Section~\ref{sec:background_selection}).  The resulting average projected number density is $\approx56$~arcmin$^{-2}$, which matches the projected number densities of our source catalogs. Fifty bright stars with apparent magnitude between 18~mag and 20~mag are simulated. In addition to fore- and backgrounds, we simulate a cluster of $\Mfiveoo=6\times10^{14}\Msun$ at $\redshift=0.42$ with the BCG in the centre and a population of early type galaxies spatially distributed following a projected NFW \citep{navarro97} profile \citep[e.g.,][]{lin04a}. We populate the cluster with galaxies between the apparent magnitudes of 18 and 25.5 according to a \citet{schechter76} luminosity function with characteristic magnitude, power law index of the faint end, and normalization measured from \citet{zenteno11}, which leads to $515$ cluster galaxies within the $R_{200}$ sphere. The half-light radius of each galaxy is randomly sampled according to the distribution of \texttt{FLUX\_RADIUS} from the source catalog extracted from the Megacam images, which is between 0\farcs15 and 1\arcsec. The half-light radius for the BCG is randomly sampled from the range  $0\farcs84\text{--}2\farcs5$, and to include the effects of saturated stars, the stellar half-light radii are randomly sampled from the range $0\farcs5\text{--}3\arcsec$.  Each object is convolved with a point spread function to reproduce the average seeing of our images. Poisson noise with the mean derived from the \rband\ data of the Megacam images is added to the images. In the end, we derive the mean of the completeness function for the source detection from these simulated images.

Figure~\ref{fig:data_depth} shows the comparison between the completeness functions of the real and the simulated data. We find that there is a good agreement for the completeness of the source detections between the simulations and the \rband\ filter, which is our detection band for cataloging. The completeness is $>94\percent$ for the background galaxies brighter than $24.0$~mag. We further derive the completeness correction as the function of the distance from the cluster centre at magnitude cut \mcut. Specifically, the completeness correction \fcom\ at \mcut\ is derived by taking the ratio of projected number density of detected galaxies between each radial bin and the radial range of $1.5\le x \le 2.5$., i.e.,
\begin{equation}
\label{eq:fcom}
\fcom (x) = \frac{n_{\mathrm{sim}}(x)}{n_{\mathrm{sim}}(1.5\le x \le 2.5)} \, ,
\end{equation}
where $x = r/\Rfiveoo$ and $n_{\mathrm{sim}}$ denotes the mean of the projected number density of the galaxies detected in the simulation (i.e., $\fcom=1$ stands for no spurious magnification bias signal created by source blending). The derived \fcom\ at $\mcut=23.5$~mag, which is the \mcut\ we use in this work (see
Section~\ref{sec:fends}), is shown in Figure~\ref{fig:fcom}. We find that the incompleteness due to blending is at level of $\approx2.5$\percent\ in the inner region of clusters ($0.1\le x \le 0.2$) and we apply this completeness correction as a function of cluster centric radius in our analysis (see Section~\ref{sec:fitting}).

%%%%%%%%%%%%%%%%
% BACKGROUND SELECTION
%%%%%%%%%%%%%%%%

%
\begin{figure*}
\centering
\includegraphics[width=\textwidth]{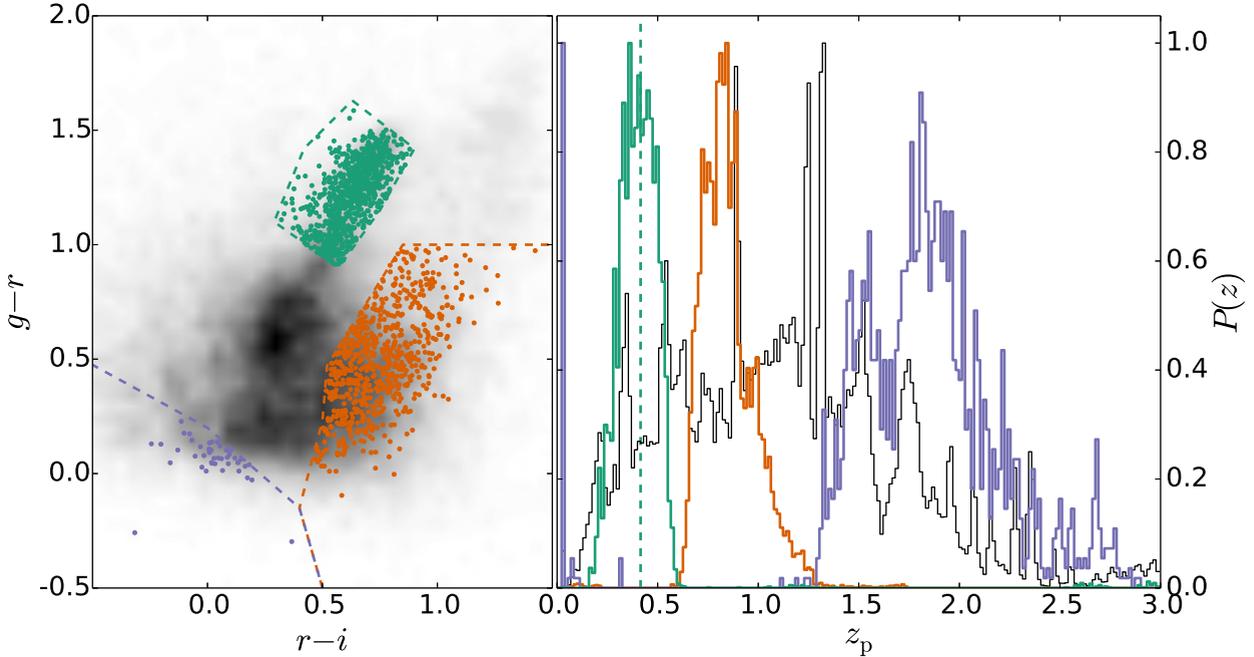}
\caption{Illustration of the colour-colour background selection in the case of SPT-CL~J0234$-$5831 ($z=0.42$) with magnitude cuts $20.0\le\gband\le23.5$. On the left is the \gr\ versus \ri\ colour-colour diagram showing the observed galaxy density distribution (gray scale), the passively evolving cluster galaxy population (green), the $z\approx0.9$ background (orange) and the $z\approx1.8$ background (blue). The corresponding normalized redshift probability distribution $P(\redshift)$ estimated from CFHTLS-DEEP for each population is shown on the right. The green dashed line marks the cluster redshift.
  \label{fig:colorcut_pz}}
\end{figure*}

\subsection{Background Selection}
\label{sec:background_selection}

Careful selection of the background galaxies is crucial for any
lensing study. It has been demonstrated that the colour selection can effectively separate galaxies at different redshifts \cite[e.g.,][]{adelberger04}.  In our case, the background galaxy population is
selected by applying colour cuts in a \gr\ versus \ri\ colour-colour space
as well as a magnitude cut in the band of interest. We first split our
cluster sample into four redshift bins from 0.25 to 0.65 in steps of
0.1 and define colour cuts corresponding to the different redshift
bins. 

The colour cut in each redshift bin is defined by three regions: a low
redshift background population, a high redshift background population,
and the passively evolving cluster galaxies at the redshift of the
bin. We define colour-colour cuts for the low- and high-\redshift\
backgrounds by tracking the colour evolution of early and late types
galaxies using the Galaxy Evolutionary Synthesis Models
\citep[\textsc{galev},][]{kotulla09}.  It has previously been shown that the low-
and high-\redshift\ backgrounds can be successfully separated from the
cluster galaxies \citep{medezinski2010}. We conservatively exclude
regions where \textsc{galev} predicts galaxy colours at the cluster
redshift for all types of galaxies.

The low-\redshift\ background is bluer (redder) than the cluster
galaxies by $\approx0.8$~mag ($\approx0.1$~mag) in $\gband - \rband$
($\rband - \iband$), while the high-\redshift\ background is bluer
than the cluster galaxies by $\approx1.2$~mag and $\approx0.6$~mag in
$\gband - \rband$ and $\rband - \iband$, respectively. By estimating
the redshift distribution of the background (see
Section~\ref{sec:lensing_effi}), the colour selection leads to the
redshift distribution of the low- and high-\redshift\ background
populations with $\left\langle\redshift\right\rangle\simeq0.9$ and
$\left\langle\redshift\right\rangle\simeq1.8$, respectively. An
example of the background selection for the redshift bin
$0.35 \leq z < 0.45$ is given in Figure~\ref{fig:colorcut_pz}.

In this work we study the magnification bias in the \gband\ band for galaxies brighter than the limiting magnitude of 23.5, given that the strongest signal for positive magnification bias is expected here (discussed further in Section~\ref{sec:fends}). We apply a magnitude cut imposing $20\leq\gband\leq23.5$ for the low and high redshift background populations selected by our colour cuts.  There are no cuts applied in the other bands.  Our final background samples provide pure background galaxy populations at low- and high-\redshift\ consistent with no cluster member contamination, as we will show in Section~\ref{sec:lensing_effi}.

%%%%%%%%%%%%%%%%
% BACKGROUND PROPERTIES
%%%%%%%%%%%%%%%%

\subsection{Background Lensing Efficiency}
\label{sec:lensing_effi}

A reliable estimate of the lensing efficiency of the background galaxies requires their redshift distribution and thus is not possible from our three band data alone. Thus, we estimate the lensing efficiency within the CFHTLS-DEEP reference field where photometric redshifts are known with a precision $\sigma_{\Delta \redshift/(1+\redshift)}=0.037$ at $\iband\le24.0$ \citep{ilbert06}.

To estimate the redshift distribution from the reference field we first select galaxies with reliable photo-$z$ estimates 
$z_{\mathrm{p}}$
by requiring $\texttt{flag\_terapix}=0$ and $\texttt{zp\_reliable}=0$ in the CFHTLS-DEEP catalog.  The cut of $\texttt{zp\_reliable}=0$ removes the galaxies due to inadequate filter coverages or problematic template fitting in the spectra energy distributions. This cut removes less than $0.25\percent$ of the galaxies in the magnitude range of interest ($\gband \le 23.5$~mag, see Section~\ref{sec:fends}); therefore, we ignore this effect. We then estimate the average lensing efficiency \lensingeff\ using the redshift distribution \Pz\ for each selected background population. Specifically, the \Pz\ for each background population is derived from the reference field with the measured photo-\redshift\ after applying the same colour and magnitude selection as in the cluster fields. Results for an example cluster are shown in the right panel of Figure~\ref{fig:colorcut_pz}, where two different background populations are identified and the passively evolving cluster population is shown for comparison. The average lensing efficiency parameter \lensingeff\ of the selected background population is estimated by averaging over the \Pz\ derived from the CFHTLS-DEEP field as
\begin{equation}
\label{eq:estimate_lensingeff}
\lensingeff_{\mathrm{t}}  = \int \Ptz \beta(\redshift, \redshift_\mathrm{l}) d\redshift \, ,
\end{equation}
where $\mathrm{t}=\{\text{low-}\redshift, \text{high-}\redshift\}$ denotes the background types and $\redshift_{\mathrm{l}}$ is the cluster redshift.

We further test the impact of  distorted redshift distributions on the estimates of \lensingeff\ for the two background populations. The redshift distribution of the background is distorted due to the fact that background galaxies at different redshifts experience different magnifications. For example, a background population with the power law index $\fends>0.4$ leads to the redshift enhancement effect \citep{coupon13} and, therefore, the average lensing efficiency deviates from the \lensingeff\ estimated from the reference field. We estimate the redshift distortion effect on our \lensingeff\ estimations as follows. We assume a background population with a power law index $\fends=0.8$ and estimate the fractional change $ {\lensingeff}_{\mathrm{l}} / \lensingeff $ in the presence of magnification caused by a cluster with $\Mfiveoo=6\times10^{14}\Msun$ at $\zd=0.42$, where
\begin{equation}
\label{eq:redshift_distortion_1}
{\lensingeff}_{\mathrm{l}} = \int P_{\mathrm{ref}}(\redshift) \magni(\Mfiveoo, \zd, \redshift)^{2.5\fends-1} \beta(z) 
\dif\redshift \, 
\end{equation}
and $P_{\mathrm{ref}}(\redshift)$ is the redshift distribution of the reference field where no lensing effect due to clusters is present.

We parametrize the cluster mass profile by the NFW model assuming the mass-concentration relation of \citet{duffy08}. This model predicts a fractional change of \lensingeff\ of at most $\approx1.6\percent$ and $\approx0.8\percent$ in the cluster inner region $0.1\le x \le 0.2$ for the low- and high-\redshift\ backgrounds, respectively. We note that the redshift distortion is more prominent for the low-\redshift\ background at $\langle\redshift\rangle\approx0.9$ because it is closer to the median redshift of our cluster sample ($\langle\zd\rangle=0.42$). Moreover, the power law index \fends\ of the low-\redshift\ background population is much lower than the assumed $\fends=0.8$ (see Section~\ref{sec:fends}). This leads us to the conclusion that the impact of redshift distortion on estimating \lensingeff\ is $<1.6\percent$. At this level, corrections for distortions of the redshift distribution to the \lensingeff\ estimations  are not needed for this analysis.
\begin{figure}
\resizebox{\hsize}{!}{\includegraphics[clip]{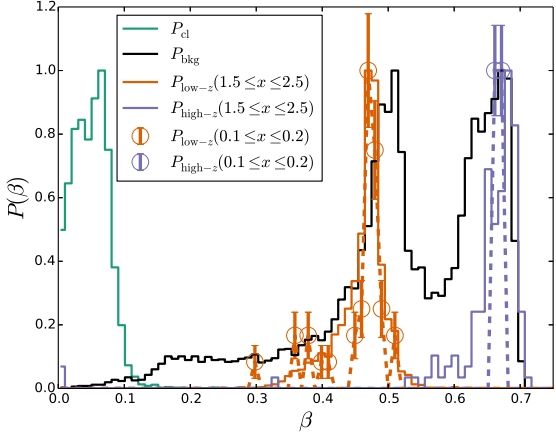}}
\caption{Comparison of the distributions of lensing efficiency \Pb\ for clusters at $0.35 \leq \zd < 0.45$. The \Pb\ for cluster galaxies (identified by $|\redshift - \zd|\le\Delta\redshift$) and the background (identified $\redshift > \zd + \Delta\redshift$) estimated from the reference field are shown in green and black solid lines, espectively. The \Pb\ estimated from the stacked low- and high-\redshift\ backgrounds are shown in orange and blue, respectively. The estimates from the outskirts ($1.5\le x \le 2.5$) and the inner core ($0.1\le x \le 0.2$) of our cluster fields are shown in solid lines and open circles, respectively, and they are in good agreement with each other and with the \Pb\ for the background determined in the reference field.  The large degree of separation between the low- and high-$z$ backgrounds and the cluster galaxies illustrates the effectiveness of colour cuts at removing cluster galaxies from the lensing source galaxy populations. Note that the tiny fraction of \Pb\ of the high-\redshift\ background at $\beta=0$ is due to the small population of the foreground galaxies instead of the cluster members (see the \Pz\ in Figure~\ref{fig:colorcut_pz}).
}
\label{fig:contam}
\end{figure}
%

%%%%%%%%%%%%%%%%
% CONTAMINATIONS
%%%%%%%%%%%%%%%%

\subsection{Cluster Member Contamination}
\label{sec:cluster_contamination}

The presence of cluster members in the selected background samples mimics the magnification signal, therefore it is crucial to quantify the cluster member contamination.
It is common in lensing studies that the reliable redshift information to separate the cluster members and background samples is not available for the observed cluster fields.  Hence, analyses often depend on information from a reference field. By leveraging a reference field, we estimate the cluster member contamination of the selected background populations by statistically connecting the observed magnitudes of the selected galaxies to the redshift information taken from the reference field. 
Specifically, we
use the method developed by \citet{gruen14}, in which they estimated the fraction of the cluster galaxies contaminating the background population by decomposing the observed distribution of the lensing efficiency, \Pb, into the known
distributions of cluster members and background galaxies. Specifically, we estimate \Pb\ of the cluster members and background from the reference field by selecting the galaxies with $|\redshift - \zd|\le \Delta\redshift$ and $\redshift > \zd + \Delta\redshift$, respectively, where \zd\ is the redshift of the cluster and $\Delta\redshift = 0.05$. 

For each galaxy $i$ with the magnitudes $\mathbf{m_{i}} = (\gband_{i},\rband_{i}, \iband_{i})$, we estimate the expected lensing efficiency $\beta(\mathbf{m_{i}})$  and the probabilities of being a cluster member and a fore/background galaxy from the galaxy sample drawn from the reference catalog within the hypersphere
$|\mathbf{m} - \mathbf{m_{i}}|\le 0.1~\mathrm{mag}$. The \Pb\ of the population is then derived from the $\beta$ estimations of the selected galaxies. We weight each galaxy by the probability of being a cluster member in deriving the \Pb\ of the cluster galaxy population, while no weight is applied in deriving the \Pb\ of the background population. 
The different magnitude distributions seen in galaxies at the cluster redshift in the cluster and in the reference fields are taken into account by applying the weighting in deriving the \Pb\ of the cluster galaxy population.
Following the same procedure, we also estimate the observed \Pb\ from the stacked background galaxies in each radial bin and in the outskirts $(1.5\le x \le 2.5)$, where $x = r / \Rfiveoo$ and \Rfiveoo\ is the cluster radius derived from the SZE-inferred mass. In this way we can decompose the observed \Pb\ and extract the fraction of the cluster galaxies contaminating the backgrounds.

The comparison of the distributions for the colour selection at $0.35\le\zd\le0.45$ is shown in Figure~\ref{fig:contam}. There is excellent agreement between the distribution of lensing efficiency in the outskirts ($1.5\le x \le 2.5$) and in the inner core ($0.1\le x \le 0.2$) regions for both low- and high-\redshift\ backgrounds. In addition, neither of them overlaps the distribution of the cluster galaxies.  The same general picture emerges for the colour selections conducted in other redshift bins. 

Following the same procedure in \cite{gruen14}, we fit the function $P_{\mathrm{m}}(\beta, x)$ to the observed distribution of $\beta$ for each radial bin to estimate the cluster contamination. Specifically, we fit the fractional cluster contamination $f_{\mathrm{cl}}(x)$ of equation~(\ref{eq:beta_decomposition}) at each radial bin $x$.
\begin{equation}
\label{eq:beta_decomposition}
P_{\mathrm{m}}(\beta, x) = f_{\mathrm{cl}}(x) P_{\mathrm{cl}}(\beta) + (1 - f_{\mathrm{cl}}(x)) P(\beta, 1.5\le x \le 2.5) \, ,
\end{equation}
where $P_{\mathrm{cl}}(\beta)$ is the distribution of $\beta$ of the cluster members estimated from the reference field and $P(\beta, 1.5\le x \le 2.5)$ is the distribution of $\beta$ of the cluster outskirt ($1.5\le x \le2.5$). We use the \cite{cash79} statistic to derive the best-fit cluster contamination $f_{\mathrm{cl}}$ and uncertainty.  Specifically, the best-fit parameters and the confidence intervals are estimated by using the likelihood estimator
\begin{equation}
\label{eq:contam_likelihood}
\begin{split}
  C_{\beta}  = 2 \sum_{i}
  \left( \vphantom{\frac{1}{2}} 
    N(x)P_{\mathrm{m}}(\beta_{i}, x) - N(\beta_{i}, x) \right.\\
  \left. + N(\beta_{i}, x) \ln
    \frac{N(\beta_{i}, x)}{
      N(x)P_{\mathrm{m}}(\beta_{i}, x)} \right) \, ,
\end{split}
\end{equation}
where $N(\beta_{i}, x)$ is the observed counts at radius $x$ for the given $\beta_{i}$ bin,  $N(x)$ is the total galaxy counts at radius $x$ (i.e., $N(x) = \sum_{i}N(\beta_{i}, x)$) and $i$ runs over the binning in $\beta$. The resulting fraction of the cluster galaxies is all zero for $x\ge 0.1$ for both backgrounds, indicating that the selected backgrounds are free from cluster galaxy contamination. We discuss the uncertainty of the measured $f_{\mathrm{cl}}$ and its impact on the mass estimates in Section~\ref{sec:systematics}.

\subsection{Power Law Index of the Galaxy Counts}
\label{sec:fends}
Estimating the power law index \fends\ (see equation~(\ref{eq:observed_n})) is crucial in magnification studies, because the magnification signal is proportional to
$\magni^{2.5\fends}$. In this analysis, we do not estimate \fends\ for each individual cluster due to the low number of background galaxies. Rather, we estimate \fends\ from the reference field with the same selection critera applied as in the cluster field. Specifically, we fit a polynomial model,
\begin{equation}
\label{eq:polymodel}
\log( N_{\mathrm{m}}(<m) ) = \frac{1}{2}am^2 + bm + c \, ,
\end{equation} 
to the observed cumulative number counts $\log(N(<m))$ brighter than magnitude $m$. In this way, the power law index at magnitude cut \mcut\ can be calculated as $\fends(\mcut) = a\mcut+b$. To estimate $\fends(\mcut)$ the fit is done locally on the interval of $-0.25 \le (m-\mcut) \le 0.25$ on binned counts with a bin width of
0.05~mag. In fitting the model we take into account the covariance among different magnitude bins in $N(<m)$; the covariance matrix is estimated by bootstrapping 2500 realizations from the catalog itself.   Specifically, the covariance matrix between magnitude bin $m_i$ and $m_j$ is built as 
\begin{equation}
\label{eq:cov_mtrx}
 C_{i,j} = 
 \left\langle 
 ( C_i - \left\langle C_i \right\rangle )
 ( C_j - \left\langle C_j \right\rangle )
\right\rangle \, ,
\end{equation}
where $C_i = \log N(<m_i)$ and the brackets $\left\langle\right\rangle$ represent an ensemble average.  The best-fit parameters of the model $(a,b,c)$ are obtained by minimizing 
\begin{equation}
\label{eq:min4s_fit}
\chi^{2} = \sum_{i,j} 
D_i \times {C^{-1}}_{i,j} \times D_j \, ,
\end{equation}
where $D_i = \log N_{\mathrm{m}}(<m_i) - \log N(<m_i)$,  $C^{-1}$ is the matrix inverse of $\left[ C_{i,j} \right]$ and $i$ and $j$ run over the ten magnitude bins in the range being fit.

\begin{figure}
\resizebox{\hsize}{!}{\includegraphics[clip]{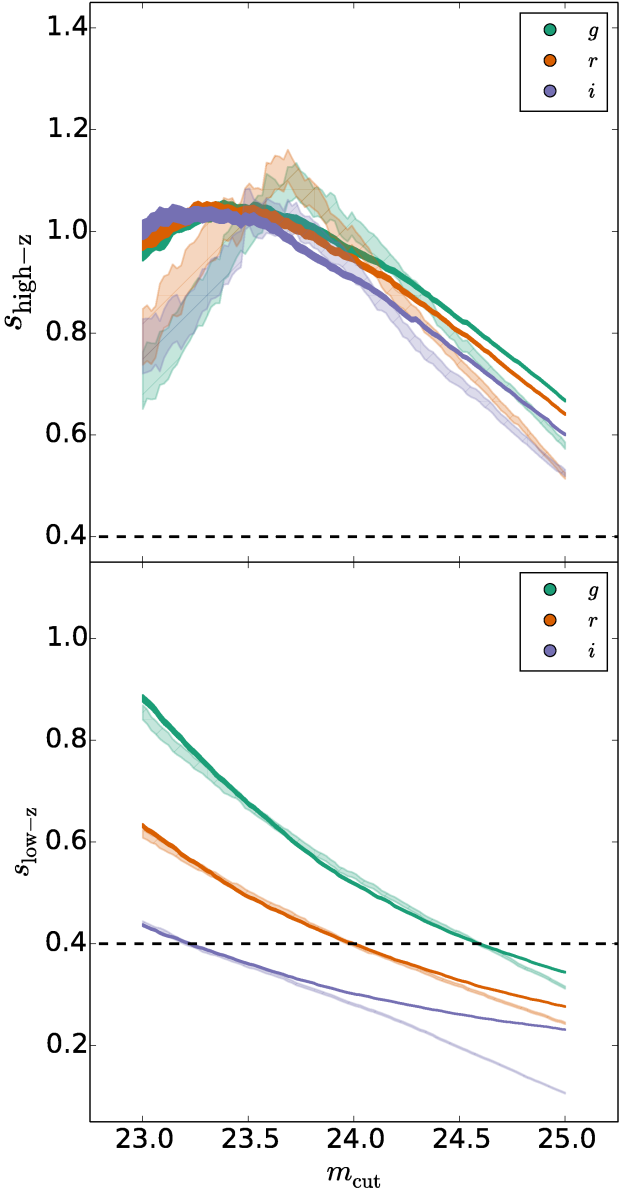}}
\vskip-0.30in
\caption{
The power law index \fends\ of the galaxy flux-magnitude distribution as a function of magnitude $m$ is shown for the high-\redshift\ population (top) and the low-\redshift\ population (bottom).  The filled and transparent regions indicate the $1\sigma$ confidence levels of the power law index \fends\ extracted from the CFHTLS-DEEP reference and the stacked SPT cluster fields, respectively. The \gband, \rband\ and \iband\ bands are colour coded in green, orange and blue, respectively. The black dashed line indicates $\fends=0.4$, where no magnification bias is expected.}
\label{fig:fends}
\end{figure}

We find that fitting this model with a range of 0.5~mag centred on the magnitude at which the slope is being measured provides an unbiased estimate of $\fends(\mcut)$ when the Poisson noise in the binned galaxy counts lies in the Gaussian regime.  Typically, we obtain $\chi^2_{\mathrm{red}}\approx1.0$ and $\chi^2_{\mathrm{red}}\approx0.8$ at $\mcut\approx23.25\text{--}24.25$ and $\mcut\approx24.25\text{--}25.0$, respectively. Furthermore, the statistical uncertainty of \fends\ is at the level of $\le1\percent$ for $23.0\le \mcut \le 25.0$. As we will discuss in Section~\ref{sec:results_and_discussion}, an uncertainty of this magnitude on \fends\ translates into a mass uncertainty of $\approx3.5\percent$,
which is small enough to have no impact on this analysis. We show the estimation of \fends\ from the reference field for the bands \gband, \rband\ and
\iband\ as a function of magnitude \mcut\ between 23~mag and 25~mag in Figure~\ref{fig:fends}, for the colour selection done in the redshift
bin between 0.35 and 0.45.

We also compare the values of \fends\ for the CFHTLS-DEEP reference field to the \fends\ measured from the cluster outskirts ($1.5\le x \le 2.5$) by stacking all 19 clusters in Figure~\ref{fig:fends}. The \fends\ estimates of the low-$z$ background show good consistency between the reference and the stacked cluster fields for \gband, \rband\ and \iband\ down to the completeness limits of our data. However, the \fends\ estimates from the stacked cluster fields tend to be lower than the ones measured from the reference field for fainter magnitudes $\mcut\geq24.0$ and in \rband\ and \iband, as one would expect given the onset of incompleteness in our dataset.  

The \fends\ measurements for the high-\redshift\ background sources from the stacked clusters do not agree as well with those from the reference fields.  For $\mcut\gtrsim23.6$~mag, the incompleteness of the high-\redshift\ background in the cluster fields starts to dominate the curvature of the source count-magnitude relation, resulting in a power law index \fends\ that is systematically smaller than the reference field.  Near $\mcut\approx23.5$ the two estimates are in agreement, but brighter than this the \fends\ is smaller in our cluster fields than in the reference fields.  This can be explained by the impact of low galaxy counts on our \fends\ estimator.  For $\mcut\lesssim23.6$~mag, the typical galaxy counts fall below 10 for the bin width of $0.05$~mag.  This leads to the  bias in the fit, which is assuming Gaussian distributed errors.  We examine this by randomly drawing 30 realizations from the reference field for the high-\redshift\ background, where each realization has the same number of galaxies as the stacked cluster field.  The bias toward low values in \fends\ from these random subsets of the reference field is consistent with that we see from the stacked cluster field, indicating that the underlying parent distributions in the cluster and reference fields are consistent.  

In summary, the high-\redshift\ background suffers more severely from low galaxy counts and incompleteness than the low-\redshift\ background (see Section~\ref{sec:background_selection}), and therefore the $\fends(\mcut)$ measurements in the stacked cluster and reference fields show better agreement.  We will discuss errors in \fends\ as a source of systematic uncertainty in Section~\ref{sec:results_and_discussion}.

To choose a magnitude cut \mcut\ that maximizes the expected magnification signal, one must consider the slope \fends\ of the count-magnitude relation, the level of Poisson noise in the lensed sample and the onset of incompleteness.  Given the depths of our photometry and the importance of the colour-colour cuts for identifying the background populations, we carry out the magnification bias analysis at $\mcut=23.5$ in \gband\ for the low- and high-\redshift\ backgrounds.  In particular, with this \gband\ cut the faintest required \iband\ magnitudes of the low- and high-$z$ population galaxies are $\approx22.3$~mag and $\approx23.5$~mag. In our data set, \iband\ is the shallowest passband, but it reaches completeness levels of $>80\percent$ at these magnitudes except in the cluster SPT-CL~J2138$-$6008.   Note that incompleteness as a function of magnitude should in principle have no effect on the derived magnification profile ($\magni^{2.5\fends - 1} = \nd(x) / \nzero(1.5\le x \le 2.5)$) as long as the incompleteness does not vary systematically with cluster radius.    At this magnitude cut \fends\ is somewhat larger than 0.75, which corresponds to an $\approx18\percent$ density enhancement for $\kappa=0.1$ assuming that $\magni \approx 1 + 2\kappa$ (see eq~\ref{eq:nd2nzero}).

%%%%%%%%%%%%%%%%
% MASKING CORRECTION
%%%%%%%%%%%%%%%%

\subsection{Masking Correction}
\label{sec:masking}

When computing object surface densities we apply a masking correction to account for regions covered by bright cluster galaxies-- mostly in the central region of the cluster-- as well as bright and extended foreground objects, saturated stars, and other observational defects. Visually identifying masked areas is not feasible for a large cluster sample and could introduce non-uniformities. We adopt the method in \cite{umetsu11} to calculate the fractional area lost to galaxies, stars and defects as a function of distance from the cluster centre.

We tune the \texttt{SExtractor} configuration parameters by setting $\texttt{DETECT\_THRESH}=5$ and $\texttt{DETECT\_MINAREA}=300$ (corresponding to 7.68~arcsec$^{2}$) to detect bright and extended objects in the coadd image and mark them in the $\texttt{CHECKIMAGE\_TYPE}=\texttt{OBJECTS}$ mode. In addition, we visually inspect the images for effects like satellite trails that typically are not captured by the \texttt{SExtractor} run. We compute the fraction of unmasked area \fmask\ where
\begin{equation}
\label{eq:def_fmask}
\fmask = \frac{A_{\mathrm{umsk}}}{A_{\mathrm{ann}}} \, ,
\end{equation}
where $A_{\mathrm{umsk}}$ is the unmasked area of the annulus and $A_{\mathrm{ann}}$ is the geometric area of the annulus.  We measure \fmask\ as a function of cluster centric distance for each cluster and use it to apply a correction to the observed density profile. On average, the unmasked fraction (see Table~\ref{tab:data_profile}) is  $\approx93\text{--}96\percent$ for all radii and greater than $\approx94\percent$ towards the cluster centre ($0.1\le x \le 0.2$). We take the masking effect into account by applying the \fmask\ correction to the fitted model in each radial bin (see Section~\ref{sec:fitting}).

%%%%%%%%%%%%%%%%
% STACKING
%%%%%%%%%%%%%%%%

\subsection{Background Profiles and Cluster Stack}
\label{sec:stacking}

We study the magnification bias of a flux-limited galaxy sample with $20.0\le\gband\le23.5$ for the low- and high-redshift background populations by stacking 19 SPT-selected clusters to enhance the signal. We stack the 19 clusters after rescaling the radii by the appropriate \Rfiveoo\ derived from the SZE-inferred masses.  This approach exploits the fact that the SZE-signature provides a low scatter mass proxy.  Given the factor of two range in mass and redshift of our sample and the availability of the SZE-inferred masses, a stack in physical radius would not be advisable.  For each of the two background populations we first derive the radial profile of the surface number density $\nd_{i}(x)$ as a function of $x=r/\Rfiveoo$ at $0.1\le x \le 2.5$ for each cluster $i$, adopting the BCG position as
the cluster centre and using the SZE derived mass to define \Rfiveoo\ (see Section~\ref{sec:sample}).
\begin{equation}
\label{eq:observable}
\nd_i(x) =  \frac{ {\Nd}_{i}(<\mcut, x) }{ {\Area}_{i}(x) {\fmask}_{i}(x) \fcom(x) } \, ,
\end{equation}
where ${\Nd}_{i}(<\mcut, x)$ is the observed cumulative number of galaxies brighter than the magnitude threshold \mcut\ that lie within a particular radial bin for the cluster and ${\Area}_{i}$ is the area of the bin. The unmasked fraction $\fmask$ is used to correct the measured galaxy counts to the full expected galaxy counts in the absence of masking. The radial correction \fcom\ is derived from our image simulations to account for the incompleteness due to blending  (see Section~\ref{sec:completeness}), and it is the same for all clusters. 

We choose bin widths of $\Delta x=0.1$ for the range $0.1\le x \le 0.5$ and $\Delta x=0.25$ at $0.5\le x \le 2.5$. The finer radial binning is used near the cluster centre because the gradient of the magnification signal is larger in the core. In the end, we stack the radial profiles to create the final stacked profile $\nd_\mathrm{tot}(x)$,
\begin{equation}
\label{eq:stack_observable}
\nd_\mathrm{tot}(x) = \sum\limits_{i=1}^{\Ncl}\, {\nd_{i}(x)}, 
\end{equation}
where $\nd_{i}(x)$ is the radial surface density profile for cluster $i$ as described above.  Note that the observed profiles are directly stacked without applying weighting. The observed magnification profile is given by
\begin{equation}
\label{eq:observed_magnifi_prfl}
\magni^{2.5\fends-1}(x) =
\frac{\nd_\mathrm{tot}(x)}{\nd_\mathrm{tot}(1.5\le x \le 2.5)} \, ,
\end{equation}
where the denominator is the mean of the counts profile in the radial range $1.5\le x\le2.5$. To compute uncertainties on the profiles, we
include Poisson noise for the galaxy number counts in each radial bin. We ignore the variance in the profiles caused by local galaxy clustering in the individual profiles because this variance is negligible compared to the Poisson noise \citep{zhang05, umetsu08, umetsu13}.  Through the stacking process both the variance due to local clustering and the Poisson noise are reduced because the cluster fields are independent.

\begin{figure}
\centering
\resizebox{\hsize}{!}{\includegraphics[clip]{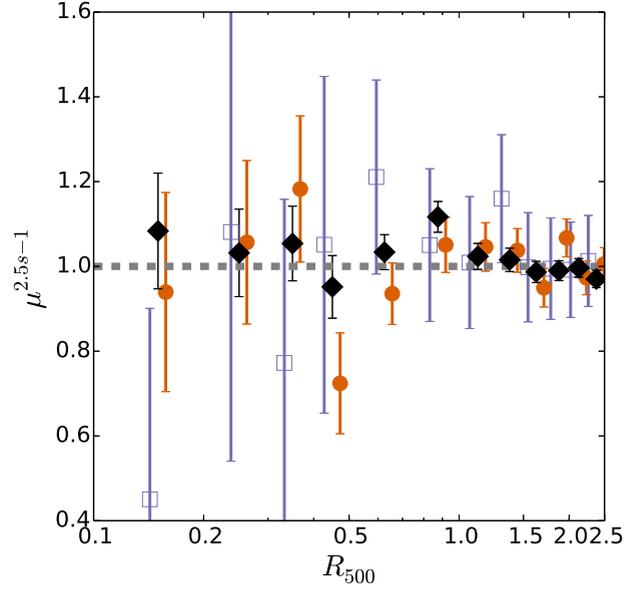}}
\caption{
 The null test on the reference field shows the normalized density profile of 20 randomly chosen apertures on the reference field after applying the same selection for the low-\redshift\ (orange circles) and high-\redshift\ (blue squares) backgrounds. The null test on the low-\redshift\ background selected in the stacked cluster field with the magnitude cut at $\rband=24$~mag (where $s=0.4$ and no net effect is expected) is shown with the black diamonds.  The red circles and blue squares are slightly offset along the horizontal axis for clarity.}
\label{fig:null_test}
\end{figure}

The same stacking procedure is performed using the reference field as a null test. Specifically, we randomly draw 20 apertures each with
\Rfiveoo\ taken to be $3\arcmin$ while avoiding any region that has been heavily masked. We stack them as in equation~(\ref{eq:stack_observable}) after applying the same background selection as for the cluster fields.  Note that the remaining masked area of the selected apertures is negligible and the procedure of stacking apertures which are randomly drawn from the reference field can remove any systematic trend of the residual masking effect. We show the resulting profiles in Figure~\ref{fig:null_test}. The variation of the  density profiles is consistent with the Poisson noise expectation and provides no evidence for an over- or under-density, providing an indication that our stacking procedure works.

After convincing ourselves that the stacking procedure on the reference field provides unbiased estimates, we then proceed to another null test on the cluster fields. This null test is defined by performing the same end-to-end analysis on the low-\redshift\ background with magnitude cut at $\rband=24$~mag instead of $\gband=23.5$~mag used in our main analysis.  The magnitude cut of $\rband=24$~mag is chosen because the low-\redshift\ background has $\fends\approx0.4$ at $\rband=24$ (see Figure~\ref{fig:fends}), and therefore we expect no magnification signal. This is a powerful end to end test of our analysis; any signal detected in this null test indicates the spurious bias in our magnification analysis. The resulting low-\redshift\ profile with the magnitude cut of $\rband=24$~mag is shown in the black diamonds in Figure~\ref{fig:null_test}.  The observed profile is consistent with $\magni=1$, and no magnification signal is seen. We hence conclude that our analysis procedure provides unbiased magnification signals.

\begin{table}
\centering
\caption{
The observed background galaxies profiles, masking correction and completeness correction.
Column~1--2: the lower and higher bound for each radial bin.
Column~3: the observed galaxy counts for the low-\redshift\ backgrounds.
Column~4: the observed galaxy counts for the high-\redshift\ backgrounds.
Column~5: the fraction of the unmasked area \fmask.
Column~6: the completeness correction \fcom\ derived from the simulation.
}
\label{tab:data_profile}
\begin{tabular}{rrrrrr}
\hline
$x_{\mathrm{lo}}$ & $x_{\mathrm{hi}}$ 
& ${\Ntot}_{, \, \mathrm{low-\redshift}}$ 
& ${\Ntot}_{, \, \mathrm{high-\redshift}}$ 
& \fmask\  &\fcom\ \\ 
\hline \hline
   $  0.10  $   & $  0.20  $   & $  35  $   & $  4  $   & $  0.953  $   &$0.979$ \\  
   $  0.20  $   & $  0.30  $   & $  34  $   & $  2  $   & $  0.948  $   &$0.977$ \\  
   $  0.30  $   & $  0.40  $   & $  50  $   & $  4  $   & $  0.946  $   &$0.987$ \\  
   $  0.40  $   & $  0.50  $   & $  66  $   & $  3  $   & $  0.942  $   &$0.997$ \\  
   $  0.50  $   & $  0.75  $   & $  224  $   & $  16  $   & $  0.932  $   &$0.996$ \\  
   $  0.75  $   & $  1.00  $   & $  326  $   & $  18  $   & $  0.948  $   &$0.998$ \\  
   $  1.00  $   & $  1.25  $   & $  352  $   & $  24  $   & $  0.931  $   &$1.000$ \\  
   $  1.25  $   & $  1.5 0 $   & $  445  $   & $  18  $   & $  0.932  $   &$0.998$ \\  
   $  1.50  $   & $  1.75  $   & $  514  $   & $  37  $   & $  0.939  $   &$0.999$ \\  
   $  1.75  $   & $  2.00  $   & $  576  $   & $  26  $   & $  0.943  $   &$0.998$ \\  
   $  2.00  $   & $  2.25  $   & $  668  $   & $  43  $   & $  0.946  $   &$1.000$ \\  
   $  2.25  $   & $  2.50  $   & $  726  $   & $  49  $   & $  0.959  $   &$1.000$ \\
\hline
\end{tabular}
\end{table}
%

%%%%%%%%%%%%%%%%
% FITTING
%%%%%%%%%%%%%%%%

\subsection{Model Fitting}
\label{sec:fitting}

To enable model fitting, we first create a stacked profile of the total observed number of galaxies $N_{\mathrm{tot}}$ above the magnitude threshold within each radial bin
\begin{equation}
\label{eq:observed_tot_gal_counts}
\Ntot(x) = \sum\limits_{i =1}^{\Ncl} \Nd_{i}(x) \, ,
\end{equation}
where ${\Nd}_i$ is the observed number of galaxies in the bin $x = r/{\Rfiveoosz}_i$ for cluster $i$ with radius ${\Rfiveoosz}_{i}$ derived using the SZE-inferred mass and the redshift.

We construct the model of the radial galaxy counts $\Ntotmodel(x)$ by stacking the predicted galaxy counts for the 19 galaxy counts models $M_{i}(x)$ using-- for each cluster $i$ at radius of $x = r/{\Rfiveoosz}_{i}$-- the average lensing efficiency ${\lensingeff}_i$, the power law index ${\fends}$, the observed background number density ${\nzero}_i$, the unmasked fraction ${\fmask}_i$ and the completeness correction $\fcom$.  Specifically, the model $\Ntotmodel(x)$ is constructed as 
\begin{equation}
\label{eq:theoretical_model1}
\Ntotmodel(x) = \sum\limits_{i =1}^{\Ncl} 
{\nmodel}_{i}(x) {\Area}_i(x) {\fmask}_{i}(x) \fcom(x) \, ,
\end{equation}
and
\begin{equation}
\label{eq:theoretical_model2}
{\nmodel}_{i}(x) = {\nzero}_{i}  
\magni({\Mfiveoo}_{i}, {\lensingeff}_{i},  x)^{2.5{\fends} - 1} \, ,
\end{equation}
where ${\nzero}_{i}$ is the number density measured in the range $1.5\le x \le 2.5$ for cluster $i$ with mass ${\Mfiveoo}_{i}$.

We parametrize the dark matter halo profile with the NFW model \citep{navarro97} assuming the mass-concentration relation of \citet{duffy08} for each cluster. During the fitting procedure we hold ${\lensingeff}_{i}$ and ${\nzero}_{i}$ for each cluster fixed at their pre-determined values, and we use the appropriate \fends\ for each of the two background populations.   We further simplify the model by fitting for a single multiplicative factor $\massfactor = {\Mfiveoo}_{i} / {\Mfiveoosz}_{i}$ for all the clusters.  Where for $\massfactor=1$ there is no net difference between the SZE-inferred and magnification masses within the full sample.  As seen in equations~(\ref{eq:theoretical_model1}) and (\ref{eq:theoretical_model2}), the model for the stacked observed galaxy counts $\Ntotmodel(x)$ is then a function of only one variable.

To estimate the best-fit mass using the observed and theoretical total galaxy number profiles $\Ntot(x)$ and $\Ntotmodel(x)$, we use the
\citet[][]{cash79} statistic. The likelihood function for fitting the magnification bias models to the total galaxy number profiles is given
by
\begin{equation}
\label{eq:fitting}
\begin{split}
  \Cstat = 2 \sum\limits_{t}\sum\limits_{j=1}^{\Nradbin}
  \left( \vphantom{\frac{1}{2}} 
    {\Ntotmodel}_{,\mathrm{t}}(\tilde{x}_j) - {\Ntot}_{,t}(x_j) \right.\\
  \left. + {\Ntot}_{,t}(x_j) \ln
    \frac{{\Ntot}_{,t}(x_j)}{
      {\Ntotmodel}_{,t}(\tilde{x}_j)} \right) \, ,
\end{split}
\end{equation}
where $t \in \{\text{low-$z$, high-$z$, combined}\}$ denotes the background populations. Note that to compare the model and the observation at the same physical radius in the space of $x = r/{\Rfiveoo}$ when $\eta\neq1$ (i.e. $\Mfiveoo\neq\Mfiveoosz$), we compare the observed profile at $x$ to the model profile at $\tilde{x}$, where $\tilde{x}=x\Rfiveoosz/\Rfiveoo=x\eta^{-{1\over3}}$.  When fitting to the combined sample, we simultaneously fit the models to the low-\redshift\ and high-\redshift\ background populations and then derive the best-fit \massfactor\ based on the sum of their \Cstat\ values (see eq~\ref{eq:fitting}).

Note that the difference of the likelihood estimator from its best-fit value $\Delta \Cstat$, is distributed like a $\chi^2$-distribution \citep[][]{cash79} and can be used to define parameter confidence intervals. Moreover, the best-fit value of $\Cstat$ can be used to test the consistency of the data and the model.

\begin{figure*}
\centering
\resizebox{\hsize}{!}{\includegraphics[clip]{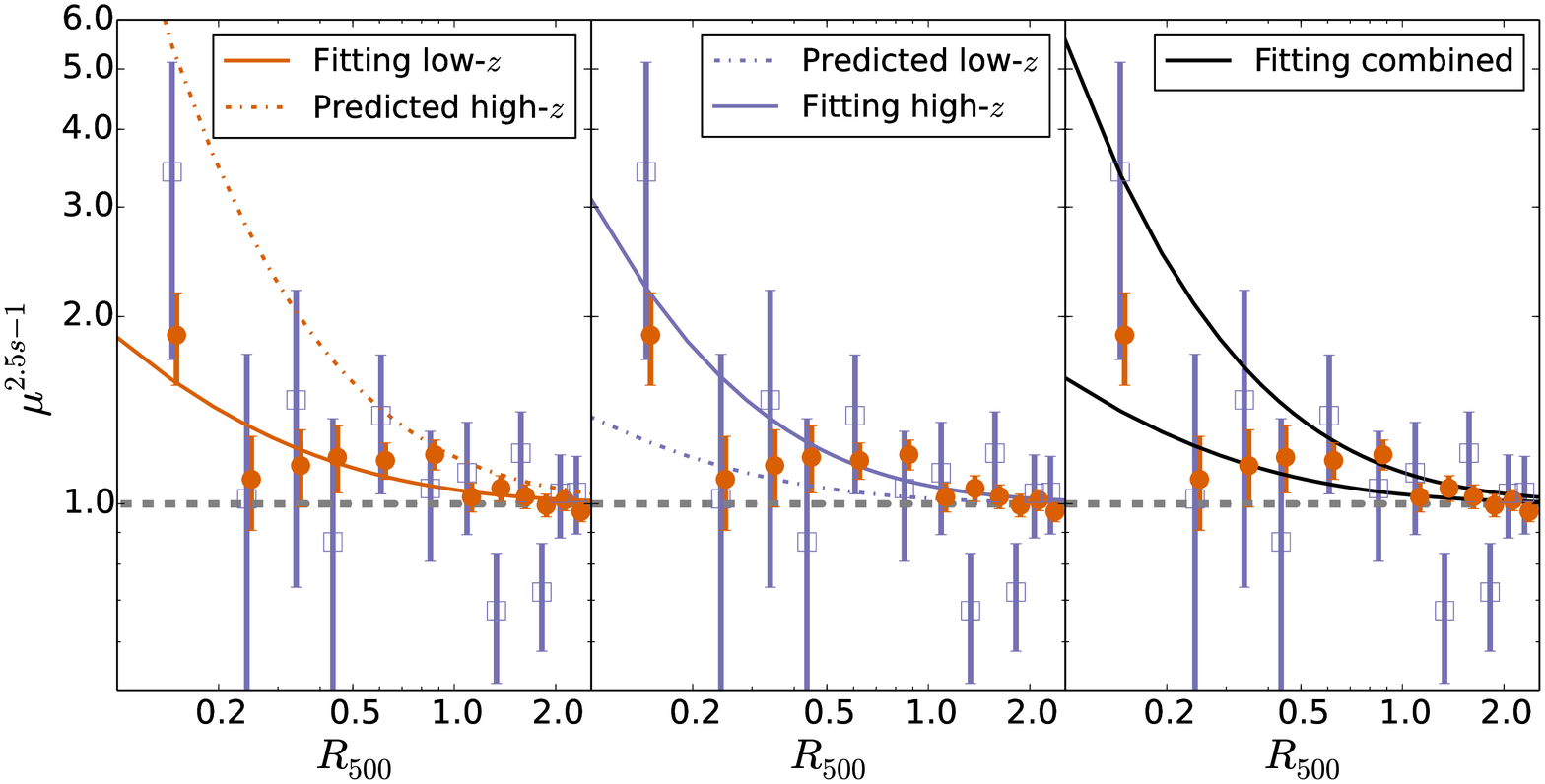}}
\vskip-0.10in
\caption{The stacked profiles for the low- and high-\redshift\ background populations with the best-fit models from different scenarios.  The panels contain the fit to the low-\redshift\ background alone (left), the fit to the high-\redshift\ population (middle), and the fit to the combined population (right).  In all three panels the orange circles (blue squares) define the stacked profile of the low-\redshift\ (high-\redshift) population, the best-fit model is defined with solid lines and the predicted profile for the other population appears as a dot-dashed line.  There is slight ($\approx1.82\sigma$) tension between the low- and high-\redshift\ populations, whereas the joint fit (right panel) is in good agreement with both populations.}
\label{fig:profiles}
\end{figure*}
\begin{table*}
\centering
\caption{
  Magnification analysis mass constraints, cross checks and detection significance.
  Column~1: background populations used in the fit.
  Column~2: best-fit \massfactor. 
  Column~3--5: 1, 2, and 3 $\sigma$ confidence level of the best-fit \massfactor.
  Column~6: reduced \Cstat\ of the fit (degree of freedom: 10, 10 and 21 for the low-\redshift, the high-\redshift\ and the combined backgrounds, respectively).
  Columns~7--8: $p$-value that the best-fit model in Column~2
  rejects the best-fit model in these columns.
  Column~9: detection significance over a model with $\massfactor=0$.%zero mass.
}
\label{tab:fitting_results}
\begin{tabular}{lcccccccc}
\hline
Population  &    \massfactor\             
&  $1\sigma_{\massfactor}$   &  $2\sigma_{\massfactor}$   &  $3\sigma_{\massfactor}$ 
& $\Cstat/\mathrm{d.o.f}$ 
& \multicolumn{2}{c}{$p$-value to reject best-fit}
& Detection Significance \\
                  &           &              &              &       
&                  & Low-\redshift                 &    High-\redshift         
&    \\
\hline
Low-\redshift\        & $1.30 $   & $^{  + 0.41 }_{ - 0.39 }$   & $^{  + 0.85 }_{ - 0.74 }$  & $^{ + 1.29 }_{ - 1.04 }$
&$1.23$ &$0.268$  &$0.075$ & $3.34\sigma$  \\ [3pt] 
High-\redshift\     & $0.46 $   & $^{  + 0.33 }_{ - 0.29 }$   & $^{  + 0.67 }_{ \cdots }$  & $^{ + 1.00 }_{ \cdots }$  
&$1.06$ &$0.061$ &$0.385$ & $1.31\sigma$  \\ [3pt] 
Combined            & $0.83 $   & $^{  + 0.24 }_{ - 0.24 }$   & $^{  + 0.49 }_{ - 0.46 }$  & $^{ + 0.74 }_{ - 0.65 }$
&$1.22$ &$0.186$ &$0.289$ & $3.57\sigma$ \\ \hline
\end{tabular}
\end{table*}
%

%%%%%%%%%%%%%%%%%%%%%%%%%%%%%%%%%%%%%%%%%%
%
% RESULTS AND DISCUSSION
%
%%%%%%%%%%%%%%%%%%%%%%%%%%%%%%%%%%%%%%%%%%

\section{Results and Discussion}
\label{sec:results_and_discussion}

We present the results in Section~\ref{sec:results} and then discuss several of the key systematic uncertainties in Section~\ref{sec:systematics}.  Section~\ref{sec:shear} contains a comparison of the observed weak lensing shear profile with the predicted profile derived from the magnification analysis.  

\subsection{Detection Significance and Mass Constraints}
\label{sec:results}

Using the procedure described in the previous section we construct the observed profiles for the ensemble of 19 massive galaxy clusters.  Properties of these profiles, including the number of background galaxies in the low- and high-\redshift\ populations and the masking and completeness corrections, are listed in Table~\ref{tab:data_profile}; the profiles themselves appear in Figure~\ref{fig:profiles}. 

We use these observed profiles to derive best-fit \massfactor\ of $\mfitlow\mfitlowonesigma$, $\mfithi\mfithionesigma$ and $\mfitcomb\mfitcombonesigma$, for the low-\redshift, high-\redshift, and combined backgrounds respectively. 
We detect the magnification bias effect at $3.3\sigma$, $1.3\sigma$ and $3.6\sigma$ for the low-\redshift, high-\redshift, and combined populations, respectively, where the confidence levels are defined via the $\Cstat$ goodness of fit statistic in comparing the observed profiles to a model with $\massfactor = 0$ (i.e. zero mass).  Table~\ref{tab:fitting_results} contains an overview of these fitting results and their statistical uncertainties.

In addition to the detection significances and confidence intervals of the best-fit masses, Table~\ref{tab:fitting_results} provides information on the statistical consistency of the best-fit models of the low-\redshift, high-\redshift, and combined background best-fit models.  The consistency between the observed profile and the best-fit model is derived using $\Cstat$.  Given the best-fit model estimated from the low-\redshift\ (high-\redshift) background population, the probability of consistency with the high-\redshift\ (low-\redshift) background population is $0.075$ ($0.061$).  In other words, there is weak tension at the $\approx1.82\sigma$ level. 

Combing both backgrounds yields  $\massfactor=\mfitcomb\mfitcombonesigma$. The probabilities of consistency of the two populations with this model are  $0.186$ and $0.289$ for the observed magnification profiles of the low- and high-\redshift\ backgrounds, respectively. 

Figure~\ref{fig:profiles} contains not only the stacked profiles for the low- and high-redshift populations but also the best-fit models.  In the left panel is the fit using only the low-\redshift\ population (solid line), but the corresponding prediction for the high-\redshift\ population is presented with the dot-dashed line.  The middle panel shows the fit to only the high-\redshift\ population (solid line) with the corresponding prediction for the low-\redshift\ population (dot-dashed line).  The right panel shows the joint fit to both populations (solid lines).  All panels contain the same observed profiles for both populations.  As is already clear from Table~\ref{tab:fitting_results}, there is weak tension between the independent fits to the low- and high-\redshift\ populations ($\approx1.82\sigma$) but the joint fit is fully consistent with both background populations.

\begin{figure}
\centering
\resizebox{\hsize}{!}{\includegraphics[clip]{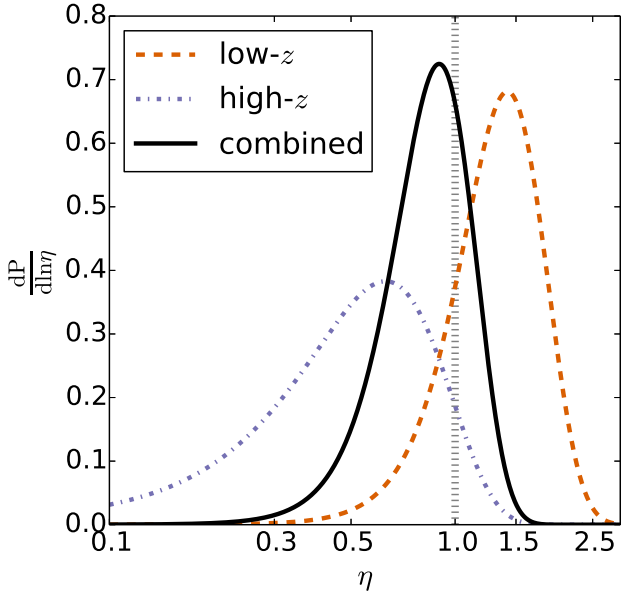}}
\vskip-0.25in
\caption{
The posterior distribution $P(\massfactor)$ of the multiplicative factor \massfactor\ given the magnification constraints. 
The grey dotted line marks $\massfactor=1$ (perfect consistency of SZE and magnification masses).  The posterior distributions $P(\massfactor)$ estimated from fitting the magnification bias model to the low-\redshift\ (orange dashed), high-\redshift\ (blue dot-dashed), and combined (black solid) populations are plotted.
None of the magnification constraints show significant tension with the  SZE-inferred masses, indicating $0.77\sigma$, $1.6\sigma$ and $0.71\sigma$ differences with $\massfactor=1$ for the low-\redshift, high-\redshift\ and combined background populations, respectively.
Note that we express the posterior distribution in $\frac{\mathrm{d}P}{\mathrm{d}\ln\massfactor} = P(\massfactor)\times\massfactor$.
}
\label{fig:eta}
\end{figure}

The posterior distributions of \massfactor\ derived by fitting the model to the low-\redshift, high-\redshift\ and combined background samples are shown in Figure~\ref{fig:eta}.  The $\massfactor=1$ (dotted line) marks the value where the SZE-inferred and magnification masses would on average be equal.  The mass factors \massfactor\ estimated from the magnification bias measurements of the low-\redshift\ (dashed line) and high-\redshift\ (dot-dashed line) backgrounds are marginally consistent with one another ($\approx1.82\sigma$ difference). The magnification constraint from the low-\redshift\ (high-\redshift) sample yields mass estimates that are $30\percent$ higher ($54\percent$ lower) than the SZE-inferred masses, corresponding to differences with $\approx0.77\sigma$ ($\approx1.6\sigma$) significance.  The combined samples prefer magnification masses that are $17\percent$ lower than the SZE-inferred masses, corresponding to a difference of $\approx0.71\sigma$.  Overall, there is no significant tension between the magnification constraints and the masses extracted using the SZE observable-mass scaling relation.

\subsection{Systematic Effects}
\label{sec:systematics}

In the following we study the influence of various sources of systematic errors on \massfactor\ including (1) differences in photometric noise in the cluster and reference fields, (2) colour biases between the two fields, (3) flux biases, (4) cluster contamination and (5) biases in the estimate of the power law slope \fends.  We explain each of these tests and the resulting impact below.  Table~\ref{tab:systematics} contains the results of our systematics tests.  

\emph{Noisy photometry $\sigma_{\mathrm{mag}}$: }To explore whether the noisier photometry in the cluster fields is impacting our analysis, we degrade the photometry of the reference field and quantify how the change of the background properties impacts the final mass factor \massfactor. Specifically, we first apply a model of magnitude uncertainty versus magnitude distribution measured from the cluster field to the reference field to degrade the photometry. We then randomly perturb the magnitude for each object in the reference field assuming the magnitudes scatter randomly following a normal distribution with a standard deviation given by the degraded magnitude uncertainty. In the end we re-measure the background properties and repeat the whole analysis to study the impact on the final best-fit \massfactor.  As can be seen in Table~\ref{tab:systematics}, the noisy photometry test results in negligible systematic uncertainties on the estimations of \lensingeff, \fends\ and  \massfactor;  this is due to the fact that the photometry noise for these bright -- relative to the completeness limit -- galaxies is small. 

\begin{figure}
\vskip-0.2in
\centering
\includegraphics[scale=0.55]{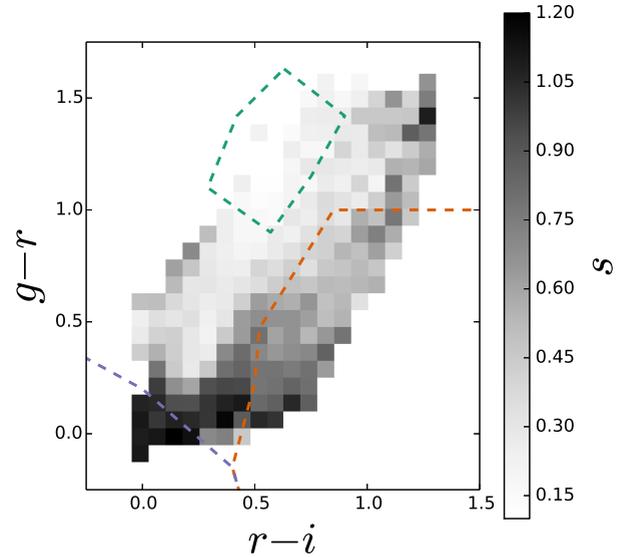}
\vskip-0.2in
\caption{
The power law index \fends\ at $\mcut=23.5$ in \gband\ band estimated from the reference field as a function of the colours ($\gband-\rband$ and $\rband - \iband$). 
The estimations of \fends\ are shown in greyscale.
The green, orange and blue dashed lines indicate the colour selections of the cluster members, the low-\redshift\ and the high-\redshift\ backgrounds, respectively, for the cluster at the redshift bin $0.35\le\redshift<0.45$.  The slope \fends\ changes rapidly with colour in the regions near the low-\redshift\ and the high-\redshift\ backgrounds.
}
\label{fig:smap}
\end{figure}

 \emph{Biased colours $\Delta~\mathrm{Colour}$:} Galaxy colour biases between the reference and cluster fields could also impact our best-fit \massfactor.  To illustrate this we measure the power law index \fends\ at $\mcut=23.5$ in \gband\ band in the reference field as a function of the colours of $\gband-\rband$ and $\rband - \iband$.    The resulting \fends-map overplotted with the colour selection of the redshift bin $0.35\le\redshift<0.45$ is shown in Figure~\ref{fig:smap}. The colour selection of the background populations can be adjusted to correspond to populations with common \fends\ and to ensure that colour boundaries do not lie where \fends\ is changing rapidly.  

We test the impact of a bias in the galaxy colours, which are calibrated with respect to the stellar locus, by shifting the whole \gr\ versus \ri\ distribution systematically by the systematic colour uncertainty 0.03~mag (see Section~\ref{sec:data}). Specifically, we shift each object in the colour-colour space by decreasing the value of \gr\ by $0.03$~mag combined with the systematics shift $\pm0.03$~mag in the colour of \ri. The objects that shift across the colour cut into the background regions are then set to have redshift zero to estimate the largest possible impact from the foreground or cluster members.  We derive the systematic uncertainties of the mass factor \massfactor\ by taking the average of the systematic mass shifts associated with the shift of $\pm0.03$~mag in \ri\ colour. We find that the slope \fends\ changes at the $\approx2\percent$ ($\approx1\percent$) level for the low-\redshift\ (high-\redshift) population, implying systematic uncertainties in \massfactor\ on the order of $\approx8\percent$ ($\approx6.7\percent$). The resulting systematic change in the combined analysis is at the level of  $\approx7\percent$.  We stress that this systematic uncertainty states the extreme case assuming all the galaxies with biased colours are cluster members. These uncertainties are smaller than the current statistical uncertainties.

\emph{Biased fluxes $\Delta~\mathrm{Flux}$:}  A bias in the absolute magnitude calibration between the cluster field and reference field could lead to a biased estimation of \fends\ for a given \mcut.  To test a flux bias at the level of the systematic flux uncertainty of $\le 0.05$~mag (see Section~\ref{sec:data}), we extract the \fends\ estimation in $\gband$ at the magnitude of $\mcut+0.05$~mag and repeat the whole analysis.  This results in a change in the estimation of \fends\ by $\approx1\percent$ in the low-\redshift\ population, leading to a shift in  \massfactor\ at the  $\approx4\percent$ level.  The impact on the high-\redshift\ population is smaller with a $\approx0.6\percent$ shift in \fends\ and a $\approx2\percent$ bias in \massfactor.

\begin{table}
\centering
\caption{
  Influence of systematic effects on the estimated \massfactor.
  Column~1: systematic.
  Column~2--3: change in  \lensingeff\ for the low- and high-\redshift\ backgrounds.
  Column~4--5: change in \fends\ for the low- and high-\redshift\ backgrounds.
  Column~6--8: change in \massfactor\ for
  fitting the low low-\redshift, high-\redshift\ and combined backgrounds.
  }
\label{tab:systematics}
\begin{tabular}{lrrrrrrr}
\hline\hline
Sources  &    \multicolumn{2}{c}{${\Delta\lensingeff\over\lensingeff}$ [\percent]} 
              &    \multicolumn{2}{c}{${\Delta\fends\over\fends}$ [\percent]} 
              &    \multicolumn{3}{c}{${\Delta\massfactor\over\massfactor}$  [\percent]}  \\
              &     Lo-$z$       &      Hi-$z$        
              &     Lo-$z$       &      Hi-$z$        
              &     Lo-$z$       &      Hi-$z$        & Cmb  \\  
\hline\hline
$\sigma_{\mathrm{mag}}$     & $ -0.1 $        & $ -1.5 $        & $ 0.3 $        & $ 1.5 $        & $ 1.2 $  & $ -0.4 $    & $ 1.0 $  \\
$\Delta$~Colour        & $ -2.4 $        & $ -4.0 $      & $ -1.9 $      & $ -0.9 $        & $ 8.0 $ & $ 6.7 $  & $ 7.4 $  \\
$\Delta$~Flux  & $\cdots$ & $\cdots$ & $ -1.0 $ & $0.6$   & $ 3.8 $ & $ -2.2 $  & $ 2.7 $  \\
Contam               & $\cdots$ & $\cdots$ & $\cdots$ & $\cdots$   & $ -2.9 $ & $ -1.7 $  & $ -2.5 $  \\
$\Delta$\fends\     & $\cdots$ & $\cdots$ & $ -0.7 $        & $ -1.6 $ & $ 3.5 $  & $ 3.0 $ & $ 3.2 $   \\
\syssig\     & $\cdots$ & $\cdots$ & $\cdots$ & $\cdots$ & $ 10.0 $  & $ 7.9 $ & $ 8.9 $   \\
\hline
\end{tabular}\\
$\dag$ \footnotesize $\Delta\equiv\mbox{(Values considering the systematics)} - \mbox{(Original values)}$. \\
\end{table}

\emph{Contamination:} In addition to studying the photometry effects, we also examine the impact of the cluster member contamination of the background populations.  The cluster contamination in the innermost bin is $0\pm0.56\percent$ inferred from the decomposition of the observed $P(\beta, 0.1\le x \le 0.2)$ of the low-\redshift\ background (see Section~\ref{sec:lensing_effi}).  The cluster contamination of the high-\redshift\ background is $0\percent$ in the inner most bin with much larger uncertainty ($\approx25\percent$) because the galaxy counts are $\approx10$ times lower than in the low-\redshift\ case. However, because the \Pb\ of the high-\redshift\ background is further separated from the \Pb\ of the cluster members than the low-\redshift\ background (see Figure~\ref{fig:contam}), the well constrained cluster contamination of the low-\redshift\ background sets a reasonable upper bound for the cluster contamination also of the high-\redshift\ population.  We therefore use the uncertainty of the cluster contamination inferred from the low-\redshift\ background as the baseline to quantify the systematic uncertainty for both populations.

We explore the impact of residual contamination by repeating the whole analysis after introducing cluster contamination into the $\Ntotmodel(x)$ with $1\percent$ contamination in the innermost bin and decreasing towards the cluster outskirts following a projected NFW profile with concentration $C_{500}=1.9$ \citep{lin04a}.  Contamination of this sort leads to a mass factor \massfactor\ biased high by $\approx3\percent$.

We further quantify the impact of cosmic variance of the derived $P_{\textrm{cl}}(\beta)$ on the estimated cluster contamination $f_{\textrm{cl}}$. Specifically, we derive the $P_{\textrm{cl}}(\beta)$ from 20 realizations, where each realization has 200 cluster galaxy members randomly drawn from the reference field. We then estimate the contribution of cosmic variance to the derived $P_{\textrm{cl}}(\beta)$ by calculating the uncertainty of the mean $P_{\textrm{cl}}(\beta)$ of these 20 realizations. As a result, the cosmic variance contributing to the derived $P_{\textrm{cl}}(\beta)$ is at the level of $\lesssim3\percent$ for a given $\beta$, indicating that the uncertainty of $f_{\textrm{cl}}$ due to cosmic variance is at the same level of $\approx3\percent$.   Accordingly, a $3\percent$ contamination would lead to a mass factor \massfactor\ biased high at the level of  $\approx9\percent$.  In this work we use the $f_{\textrm{cl}}(0.1\le x \le 0.2)=1\percent$, which is $\approx2$ times of the derived statistical uncertainty of $f_{\textrm{cl}}$, to estimate the systematic uncertainty of  \massfactor.  The resulting change in mass estimates is shown in Table~\ref{tab:systematics}.
We stress that the proper uncertainty of cluster contamination $f_{\textrm{cl}}$ estimated from the method of \cite{gruen14} should include both the statistical variation of the observed \Pb\ at each radial bin and the cosmic variance of the derived $P_{\textrm{cl}}(\beta)$ of cluster members. In this work, we only use the statistical uncertainty of the radial fitting while fixing the derived $P_{\textrm{cl}}(\beta)$ and $P(\beta, 1.5 \le x \le 2.5)$.

\emph{Biased slope $\Delta$\fends:}  We quantify the systematic uncertainty (see Section~\ref{sec:fends})
caused by the differences between the values of \fends\ measured in the cluster and reference fields.  The difference of the measured \fends\ between the reference and cluster fields is negligible, causing the systematic uncertainties of  \massfactor\ at the level  $\lesssim 3.5\percent$ for fitting the low-\redshift,
high-\redshift\ and combined backgrounds. 

\begin{figure*}
\centering
\includegraphics[width=\textwidth]{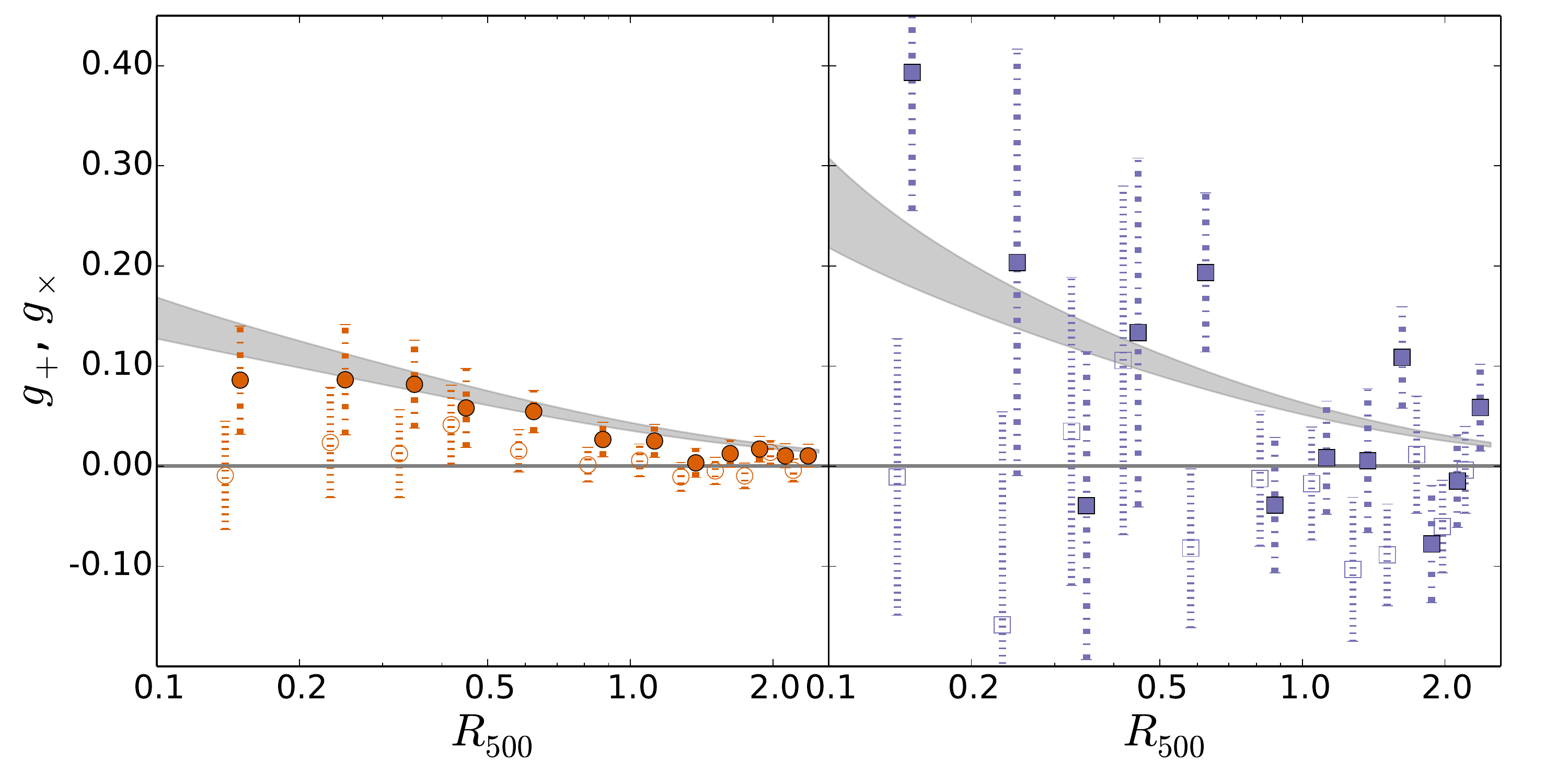}
\vskip-0.15in
\caption{ The shear profiles from the stacked catalogs with the model prediction based on the best-fit  \massfactor\ of the  magnification analysis. The \massfactor\ estimated from fitting to the combined background populations yields a mass estimate of $(5.37\pm1.56)\times10^{14}\Msun$ given the mean of 19 SZE-inferred masses.  The shear profile of the low-redshift background is shown in the left panel, and that of the high-redshift background is shown in the right panel. The open and solid points indicate the tangential shear (\sheartan) and cross shear (\shearcro) components of the reduced shear, respectively.  The gray shaded regions show the shear profile predictions with $1\sigma$ confidence region.  Data points are horizontally offset for clarity.}
\label{fig:shear}
\end{figure*}

\emph{Total systematic uncertainty \syssig:} Reviewing these estimates, the most important source of systematic uncertainty in the best-fit \massfactor\ comes from colour biases.  Thereafter, the other sources aside from noisy photometry are of roughly equal importance.  It is important to note that if the cluster contamination cannot be controlled at the level of $\approx1\percent$ then it could easily become the dominant source of systematic uncertainty.  
The systematic uncertainties of the mass are normally distributed
in the limit of small perturbations seen in the background properties (\lensingeff\ and \fends).  We therefore combine these estimates in quadrature.
The total estimated systematic uncertainties for the mass of the low- and high-\redshift\ populations and the combined analysis are $\syssig=10.0\percent$, $\syssig=7.9\percent$ and $\syssig=8.9\percent$, respectively.  This results in the final mass factor \massfactor\ of $\mfitlow\mfitlowonesigma\mathrm{(stat)}\mfitlowsys\mathrm{(sys)}$,  $\mfithi\mfithionesigma\mathrm{(stat)}\mfithisys\mathrm{(sys)}$ and $\mfitcomb\mfitcombonesigma\mathrm{(stat)}\mfitcombsys\mathrm{(sys)}$
from the analysis of the low-\redshift, high-\redshift, and combined background populations, respectively.  The statistical uncertainties are dominant in all cases.

\subsection{Comparison to Shear Profile}
\label{sec:shear}

As a cross-check we examine whether the weak lensing shear profiles agree with the expectation, given our weak lensing magnification constraints.  To construct the shear profiles we use the shape catalogs derived as described elsewhere \citep[][Dietrich et al. in prep.]{hoekstra12,hoekstra15} of the low-\redshift\ and high-\redshift\ populations with exactly the same colour selections used in our magnification analysis.  We stack the shape catalogs after the colour and magnitude selections. We find that $5.2\percent$ ($3.3\percent$) of the low-\redshift\ (high-\redshift) background galaxies do not have shape measurements, which is mainly due to blending issues associated with the different source finder (i.e. \texttt{hfindpeaks}) used in the shape measurement pipeline).  We stress that the shear profile is less sensitive to the missing objects due to blending than the magnification analysis, we hence ignore this effect in deriving stacked shear profiles.
We derive both tangential shear (\sheartan) and cross shear (\shearcro) profiles
including the lensing weight \citep[][Dietrich et al. in prep.]{hoekstra12,hoekstra15} of each single galaxy.
We predict the \sheartan\ profile using the best-fit  \massfactor, the average lensing efficiency \lensingeff\ for each background population, and a fiducial cluster located at the median redshift of the 19 clusters.  Specifically, we use the mean of the 19 SZE-inferred masses ($6.47\times10^{14}\Msun$) multiplied by the \massfactor\ estimated from fitting the combined background populations, which is consistent with the SZE-inferred masses, as the input mass for predicting the shear profiles.  As a result, the \massfactor\ estimated from fitting the combined background samples yields a mass estimate of $(5.37\pm1.56)\times 10^{14}\Msun$.

Figure~\ref{fig:shear} shows the predicted and observed shear profiles for the low- and high-\redshift\ populations. We emphasize that the gray shaded regions are model shear profiles derived from the magnification analysis and not fits to the observed shear profiles.  Both low- and high-\redshift\ population observed shear profiles are statistically consistent with the predicted shear profiles. The probabilities that the data are described by the model are $0.87$ and $0.25$, for the low- and high-\redshift\ backgrounds, respectively.  The observed cross shear (\shearcro) profiles are both consistent with zero. The good consistency between the observed tangential shear (\sheartan) profiles and the models inferred from the magnification analysis provides a clear indication that the magnification bias signal we observe is not a spurious signal caused by cluster members.  Such contaminating cluster member galaxies would not have a tangential alignment with respect to the cluster centre.

%%%%%%%%%%%%%%%%%%%%%%%%%%%%%%%%%%%%%%%%%%
%
% CONCLUSION
%
%%%%%%%%%%%%%%%%%%%%%%%%%%%%%%%%%%%%%%%%%%

\section{Conclusions}
\label{sec:conclusion}

By stacking the signal from 19 massive clusters with a mean SZE-inferred mass of  $\Mfiveoo=(6.47\pm0.31)\times10^{14}\Msun$, we detect the enhancement in the number density of a
flux-limited ($\gband\le23.5$) and colour (\gr\ and \ri) selected background population with $\redshift\approx0.9$ at $3.3\sigma$ confidence.  We find only very weak indications of the magnification bias signal using the same data but colour selecting for a higher redshift background population at $\redshift\approx1.8$. This background sample at $\redshift\approx1.8$ is significantly smaller than that at $\redshift\approx0.9$, increasing the Poisson noise and thus reducing the significance of the measurement.

We estimate the  mass factor \massfactor, which is the ratio of the magnification and SZE-inferred masses extracted from the whole cluster ensemble. We find a best-fit \massfactor\ of $\mfitcomb\mfitcombonesigma\mathrm{(stat)}\mfitcombsys\mathrm{(sys)}$ by fitting to the combined low- and high-redshift background populations. This indicates that there is no tension between the magnification masses and those estimated using the SZE observable-mass relation.

We analyze the potential impact of systematic errors caused by photometric scatter and bias, cluster galaxy contamination, and uncertainties in the estimation of the average lensing efficiency \lensingeff\ and power law index \fends\ of the galaxy count-magnitude relation for each of the two background populations. We quantify how these effects impact the final mass factor \massfactor\ estimated from the fit.  We find that the systematic source with the largest potential to affect \massfactor\ estimates ($7.4\percent$ bias for the combined constraint) is the bias in the calibration of the photometric colour, which would lead to an inconsistency between the estimation of the background properties of the data and the reference field.   The other biases each contribute systematic uncertainties at the $\le5\percent$ level, which we combine in quadrature to estimate a final $7.4\percent$ systematic uncertainty on the combined \massfactor\ constraint.  We conclude that the mass constraints can be reliably estimated using the magnification bias if the unbiased background properties can be estimated.  The uncertainty of the \massfactor\ estimation in this work is currently dominated by the statistical uncertainty, which is due to the lack of background galaxies needed to suppress the Poisson noise.

This work underscores the promise of using magnification bias of normal background galaxy populations to calibrate the observable-mass scaling relation and measure cluster masses in multi-band imaging survey data with depths similar to those in the Dark Energy Survey.  For the clusters detected in the 2500 $\deg^{2}$ SPT-SZ survey, there are $\approx200$ with redshifts $0.3\le \redshift \le 0.6$.  By carrying out the same analysis as discussed in this work, we expect the detection significance of the magnification effect would be increased to $\approx10\sigma$ and $\approx4\sigma$ for the low-\redshift\ and high-\redshift\ background populations, respectively.  Therefore, by stacking samples of hundreds to thousands of clusters in such a dataset, it is possible to carry out important cross-checks of masses extracted through weak lensing shear, galaxy dynamics and other methods.

%%%%%%%%%%%%%%%%%%%%%%%%%%%%%%%%%%%%%%%%%%
%
% ACKNOWLEDGE
%
%%%%%%%%%%%%%%%%%%%%%%%%%%%%%%%%%%%%%%%%%%

\section*{Acknowledgements}
\label{sec:acknowledgements}
We acknowledge the support by the DFG Cluster of Excellence ``Origin and Structure of the Universe'' and the Transregio program TR33 ``The Dark Universe''. D.A. and T.S. acknowledge support from the German Federal Ministry of Economics and Technology (BMWi) provided through DLR under projects 50 OR 1210, 50 OR 1308, and 50 OR 1407. CR acknowledges support from the Australian Research Council's Discovery Projects scheme (DP150103208). The South Pole Telescope is supported by the National Science Foundation through grant ANT-0638937. Partial support is also provided by the NSF Physics Frontier Center grant PHY-0114422 to the Kavli Institute of Cosmological Physics at the University of Chicago, the Kavli Foundation and the Gordon and Betty Moore Foundation.   Optical imaging data were obtained with Megacam on the 6.5~m Magellan Clay Telescope.

Facilities:  South Pole Telescope, Magellan

%%%%%%%%%%%%%%%%%%%%%%%%%%%%%%%%%%%%%%%%%%
%
% Bibliography
%
%%%%%%%%%%%%%%%%%%%%%%%%%%%%%%%%%%%%%%%%%%

\bibliographystyle{mn2e}
\bibliography{spt}

\end{document}